\documentclass[11pt]{article}
\usepackage{axodraw}
\usepackage{epsfig}
\usepackage{amsfonts}
\usepackage{amsmath}
\usepackage{bbm}
\usepackage{cite}
\allowdisplaybreaks[1]

\input pix.sty

\newcommand{\Epsilon}{\mathcal{E}}
\newcommand{\Xs}{{X}^{ }_1}
\newcommand{\Xo}{{X}^{ }_8}
\newcommand{\Ss}{{S}^{ }_1}
\newcommand{\So}{{S}^{ }_8}

\renewcommand{\eq}{eq.~}
\renewcommand{\eqs}{eqs.~}
\newcommand{\Eqs}{Eqs.~}
\renewcommand{\se}{sec.~}
\renewcommand{\ses}{secs.~}
\renewcommand{\fig}{fig.~}
\renewcommand{\figs}{figs.~}

\newcommand{\tinymsbar}{{\overline{\mbox{\tiny\rm{MS}}}}}
\newcommand{\Lambdamsbar}{{\Lambda_\tinymsbar}}

\newcommand{\alphas}{\alpha_{\rm s}}
\newcommand{\Nf}{N_{\rm f}}
\newcommand{\Nc}{N_{\rm c}}

\newcommand{\Nt}{N_\tau}
\newcommand{\Tc}{T_{\rm c}}

\newcommand{\rmO}{{\mathcal{O}}}

\newcommand{\CF}{C_\rmii{F}}

\def\lsi{\raise0.3ex\hbox{$<$\kern-0.75em\raise-1.1ex\hbox{$\sim$}}}
\def\gsi{\raise0.3ex\hbox{$>$\kern-0.75em\raise-1.1ex\hbox{$\sim$}}}
\newcommand{\lsim}{\mathop{\lsi}}
\newcommand{\gsim}{\mathop{\gsi}}

\newcommand{\nB}{n_\rmii{B}}

 \renewcommand{\nB}[1]{f_\rmii{B{#1}}}

\newcommand{\rmii}[1]{{\mbox{\tiny\rm{#1}}}}

\newcommand{\re}{\mathop{\mbox{Re}}}
\newcommand{\im}{\mathop{\mbox{Im}}}

\newcommand{\Tint}[1]{{\hbox{$\sum$}\!\!\!\!\!\!\!\int\,}_{\!\!\!\!\raise-0.9ex\hbox{$\scriptstyle{#1}$}}}
\newcommand{\Tinti}[1]{{{\Sigma}\!\!\!\!\raise0.3ex\hbox{$\int$}_\rmii{${#1}$}}}

\newcommand{\unit}{{\mathbbm{1}}} 
\newcommand{\bi}{\begin{itemize}}
\newcommand{\ei}{\end{itemize}}
\newcommand{\hide}[1]{ }

\newcommand{\ff}{\rmi{\sl f\,}}

\newcommand{\deltabar}{\delta\!\!\!\raise0.7ex\hbox{--}\,}
\def\TAsc(#1,#2)(#3,#4,#5)%
{\SetWidth{2.0}\CArc(#1,#2)(#3,#4,#5)\SetWidth{1.0}}
\def\Lwidth{3}

\def\TAgl(#1,#2)(#3,#4,#5){\SetWidth{2.0}\PhotonArc(#1,#2)(#3,#4,#5){\Lwidth}%
{6.283 #3 mul 360 div #4 #5 sub #4 #5 sub mul sqrt mul Tdensity mul}%
\SetWidth{1.0}}
\def\TLgl(#1,#2)(#3,#4){\SetWidth{2.0}\Photon(#1,#2)(#3,#4){\Lwidth}
{#1 #3 sub #1 #3 sub mul #2 #4 sub #2 #4 sub mul add sqrt Tdensity mul}%
\SetWidth{1.0}}
\newcommand{\piC}[1]{\;\parbox[c]{120pt}{\begin{picture}(120,60)(0,0)
\SetWidth{1.0}\SetScale{1.2} #1 \end{picture}}\; }
\renewcommand{\picc}[1]{\;\parbox[c]{60pt}{\begin{picture}(60,30)(0,0)
\SetWidth{1.0}\SetScale{1.3} #1 \end{picture}}\; }
\def\Lwidth{1.3}

\def\Gcirc{\piC{%
 \SetWidth{1.5} 
 \CArc(60,30)(25,-89,269)%
 \SetWidth{1.0}
 \DashCArc(60,30)(22,45,315){1}%
 \ArrowArc(60,30)(25,255,270)%
 \Line(61,2.2)(61,8)%
 \Line(75,15)(81,9)%
 \Line(75,45)(81,51)%
 \Text(75,-2)[l]{$0$}%
 \Text(71,-2)[r]{$\beta$}%
 \Text(101,5)[c]{$\tau_2$}%
 \Text(102,65)[c]{$\tau_1$}%
 \Text(48,38)[l]{$G^\theta_{ }$}%
}}
\makeatletter \@addtoreset{equation}{section} \makeatother
\renewcommand{\theequation}{\arabic{section}.\arabic{equation}}
\makeatletter
\renewcommand\section{\@startsection {section}{1}{\z@}%
                                   {-5.5ex \@plus -1ex \@minus -.2ex}
                                   {2.3ex \@plus.2ex}%
                                   {\normalfont\large\bfseries}}
\renewcommand\subsection{\@startsection{subsection}{2}{\z@}%
                                     {-3.25ex\@plus -1ex \@minus -.2ex}%
                                     {1.5ex \@plus .2ex}%
                                     {\normalfont\normalsize\bfseries}}
\renewcommand\thesection {\@arabic\c@section}
\renewcommand\thesubsection   {\thesection.\@arabic\c@subsection}
\renewcommand{\@seccntformat}[1]{%
\csname the#1\endcsname.\hspace{1.0em}}
\makeatother

\begin{document}

\flushbottom

\begin{titlepage}

\begin{flushright}
June 2016
\vspace*{1cm}
\end{flushright} 
\begin{centering}

\vfill

{\Large{\bf
 Rapid thermal co-annihilation through bound states in QCD 
}} 

\vspace{0.8cm}

Seyong Kim$^{\rm a}$ and 
M.~Laine$^{\rm b}$ 

\vspace{0.8cm}

$^\rmi{a}$%
{\em
Department of Physics, 
Sejong University, Seoul 143-747, South Korea\\}

\vspace*{0.3cm}

$^\rmi{b}$%
{\em
AEC, Institute for Theoretical Physics, 
University of Bern, \\ 
Sidlerstrasse 5, CH-3012 Bern, Switzerland\\} 

\vspace*{0.8cm}

\mbox{\bf Abstract}

\end{centering}

\vspace*{0.3cm}
 
\noindent
The co-annihilation rate of heavy particles close to thermal equilibrium,
which plays a role in many classic dark matter scenarios, can be ``simulated''
in QCD by considering the pair annihilation rate of a heavy quark and
antiquark at a temperature of a few hundred MeV. We show that the so-called
Sommerfeld factors, parameterizing the rate, can be defined and measured
non-perturbatively within the NRQCD framework. Lattice measurements indicate a
modest suppression in the octet channel, in reasonable agreement with
perturbation theory, and a large enhancement in the singlet channel, much
above the perturbative prediction. The additional enhancement is suggested
to originate from bound state formation and subsequent decay. Making use of
a Green's function based method to incorporate thermal 
corrections in perturbative co-annihilation rate computations, we show
that qualitative agreement with lattice data can be found once 
thermally broadened bound states are accounted for. We suggest that
our formalism may also be applicable to specific dark matter models which
have complicated bound state structures.

\vfill

 

\vfill

\end{titlepage}

%
\section{Introduction}

Heavy particle co-annihilation represents a subtle
problem, because the slowly moving particles experience 
strong initial state effects before the final inelastic process.
Reversing the time direction, the same can be said about
pair creation. Nevertheless, within two-body quantum mechanics 
in a system with a Coulomb potential, this physics 
was understood long ago~\cite{asommerfeld,landau3}. 
It also plays an essential role 
for heavy particle pair creation in QCD~\cite{fadin}. The enhancement
or suppression of the annihilation rate, depending on whether the
Coulomb interaction is attractive or repulsive, is generically 
referred to as the Sommerfeld effect. 

If the co-annihilating particles are part of a thermal medium, 
the Sommerfeld effect plays a role in a modified form. 
In this case the velocities of the annihilating particles are not 
fixed, but come from a statistical distribution. In particular, 
if we assume the heavy particles to be in kinetic equilibrium, 
the velocities are distributed according to the Boltzmann weight. 
Then we speak of thermally averaged Sommerfeld factors. The thermally
averaged Sommerfeld factors are relevant for cosmology, 
particularly for determining the abundance of heavy weakly 
interacting dark matter particle species~\cite{hisano,sfeldx,feng,iengo}, 
which decouple deep in the non-relativistic regime. A recent example of
an embedding of the corresponding Sommerfeld factors 
in a realistic setting can be found in ref.~\cite{sfeldy}.  

The starting point of the present paper is
the observation that physics formally similar to 
dark matter co-annihilation takes place with heavy quarks and antiquarks
in hot QCD at a temperature of a few hundred MeV. Going beyond 
a leading-order computation based on Boltzmann equations~\cite{s1,s2,s3}, 
the rate of their chemical equilibration can be defined on a general
level~\cite{chemical}, and can be shown to be sensitive to the 
physics of the thermal Sommerfeld effect~\cite{pert}. 
Even though the associated time scale appears to be too long
to play a practical role within the life-time $\sim 10$~fm/c of a 
fireball generated in a heavy-ion collision experiment, this analogy
nevertheless means that established methods of QCD, 
including those of lattice QCD, 
can be used to investigate the problem. In particular, in 
ref.~\cite{pert} a strategy was outlined for implementing 
in Euclidean spacetime the absorptive parts of 4-quark operators,
which can be used for describing the decays of quark-antiquark states
within the NRQCD effective theory framework~\cite{nrqcd,bodwin}. 

The purpose of the present paper is to realize this proposal. 
A main part is to develop further the 
theoretical formulation, showing
that the measurement can be reduced to 2-point functions 
and that analytic continuation back to Minkowskian spacetime 
poses no problem. We also carry out an 
exploratory lattice study, finding
an intriguing pattern with an enhancement in the singlet channel
much larger than predicted by the standard formulae used in the 
literature for incorporating thermal Sommerfeld enhancement. 
An explanation for this observation in terms of bound-state
physics is put forward. 

The plan of this paper is the following. After defining the 
basic observables in \se\ref{se:defs}, we show in~\se\ref{se:theory}, 
through a spectral representation and a canonical analysis, 
how their measurement can be related to purely static 2-point
correlation function ratios. The perturbative evaluation of
these correlators is discussed in \se\ref{se:pert}, and
corresponding lattice measurements are reported 
in~\se\ref{se:lattice}. A concluding discussion and an outlook
are offered in \se\ref{se:concl}. 
Readers not interested in the details of the theoretical and 
lattice analyses are suggested to consult \se\ref{ss:physics}
and subsequently proceed to \ses\ref{se:pert} and \ref{se:concl}.

%
\section{Setup} 
\la{se:defs}

%
\subsection{Physics background} 
\la{ss:physics}

In order to outline the physics problem in a simple 
setting, we consider heavy quarks and antiquarks, 
with a pole mass $M$,
placed in a heat bath at a temperature~$T \ll M$. 
We stress that even though we make use of QCD terminology, 
the discussion until \eq\nr{spinor} applies rather generally, with the
gauge group replaced as appropriate. 
In the QCD context the only specific assumption made is that 
we consider time scales up to some thousands of fm/c, so that
the weak decays of the heavy quarks can be omitted; this can 
be viewed as the analogue of ``R-parity conservation'' in 
supersymmetric theories. 
Thereby the only number-changing reactions are 
pair annihilations and creations. 

In the heavy-quark limit, QCD has an ``emergent'' 
symmetry, in that the quark and antiquark numbers are conserved 
separately. This conservation is only violated by higher-dimensional
operators suppressed by $1/M$. Omitting such operators
it is possible to define separate 
distribution functions for heavy quarks and antiquarks. 
We assume that both are 
close to kinetic and chemical equilibrium, 
whereby the distribution functions have the forms 
$f_p = \bar{f}_p \approx 2\Nc \exp(- E_p/T)$, where 
$E_p \equiv \sqrt{p^2 + M^2}$ and $2\Nc$ counts the spin and 
colour degrees of freedom. 
Here $p\equiv |\vc{p}|$ is the spatial
momentum with respect to the heat bath. 

Consider now the total number density, 
$n = \int_\vc{p} (f_p + \bar{f}_p)+ n_\rmi{bound}$, 
where $n_\rmi{bound}$ denotes the density of quark-antiquark
bound states. 
The equilibrium value of the total density reads 
$n_\rmi{eq} \approx 4 \Nc \int_\vc{p} \exp(- E_p/T) + 
\rmO(e^{-2M/T})$, where the last term indicates the 
bound state contribution. 
If the equilibrium
conditions are evolving, for instance through a Hubble expansion 
characterized
by a rate $H$, then the system attempts to adjust its 
number density to this change. Within a Boltzmann equation approach
the evolution equation has the form~\cite{bbf}
\be
 (\partial_t + 3 H )\, n \approx - c\, (n^2 - n_\rmi{eq}^2) 
 \;. \la{Hubble}
\ee
If we linearize the right-hand side around equilibrium, this can be 
rewritten as 
\be
 (\partial_t + 3 H )\, n = - \Gamma_\rmi{chem}(n- n_\rmi{eq}) + 
 \rmO(n - n_\rmi{eq})^2
 \;. \la{Boltzmann}
\ee
The coefficient $\Gamma_\rmi{chem} = 2\,  c\, n_\rmi{eq}$
is called the {\em chemical equilibration rate}. It tells how
efficiently the system is able to re-adjust its density towards
the evolving $n_\rmi{eq}$, and encodes the effects of
the microscopic processes which can change the quark and antiquark
number densities, notably pair creations and annihilations. 
Note that, unlike \eq\nr{Hubble}, the form of \eq\nr{Boltzmann}
represents a general linearization 
and is thus valid beyond the Boltzmann approach. 

Let us stress that $ \Gamma_\rmi{chem} $ describes the {\em slow} evolution
of a number density, averaged over a large volume. The associated
frequency scale is  $\omega \sim \Gamma_\rmi{chem} \sim T \exp(-M/T) \ll T$.
This slow evolution is 
caused by infinitely many rare individual processes. The individual 
processes are associated with pair creations and annihilations and 
carry a {\em large} energy, $E \sim 2 M \gg T$. 
We are not studying these individual reactions separately; 
in fact the energy scale $2M$ can be integrated out whereby the 
system can be described by an effective
theory known as the NRQCD~\cite{nrqcd,bodwin}, 
capturing physics at scales $|E - 2 M| \ll M$.

Now, it is important to realize that even though bound states give
a strongly suppressed contribution to $n_\rmi{eq}$, they give
an enhanced contribution to $\Gamma_\rmi{chem}$. 
Indeed, as already mentioned, 
in $n_\rmi{eq} \sim e^{-M/T}$ the bound state contribution
is $n_\rmi{bound} \sim e^{-2M/T}$. But $\Gamma_\rmi{chem}$
originates from reactions where a quark and antiquark come together, 
i.e.\ $|\partial_t n| \sim  e^{-2M/T}$. In a bound state the
quark and antiquark are ``already'' together, and with
a less suppressed Boltzmann weight, because of a binding energy
$\Delta E > 0$. Therefore the decay rate from bound states is 
\be
 |\partial_t n_\rmi{bound}| \sim  e^{-(2 M-\Delta E)/T }
 \;.  \la{robust}
\ee
If $T \lsim \Delta E \sim M \alphas^2$, 
this contribution dominates the annihilation rate. 
This leads to a physical 
picture in which chemical equilibration proceeds via a ``two-stage''
process: in thermal equilibrium, only a small fraction of quarks and 
antiquarks form bound states. But it is those bound states which
are most efficiently depleted as the temperature decreases. 
Bound state formation itself is a fast process, without 
any exponential suppression factors. 

Even though the potential significance of bound states has been 
mentioned in many cosmological 
studies, such as refs.~\cite{old1,old2}, they have not established
themselves as a standard ingredient in quantitative estimates
of the thermal freeze-out process. 
One good reason is certainly that bound states only exist in very specific 
weakly interacting models. But there is also another issue, namely
that the computations are usually based on the Boltzmann
equation, \eq\nr{Hubble}, where $c = \langle \sigma v \rangle$
is a thermally averaged annihilation cross section. 
Thereby, if not sufficient care is taken, the basic assumption
that the system can be described by dilute
single-particle on-shell states may be inadvertently built into
the formalism from the beginning. Recently the importance
of bound-state effects has been stressed in, e.g., 
refs.~\cite{old3,old35,old4,old45,old5}, which also suggested 
various ways to include them.  
The simplest possibility is to add a bound state phase space 
distribution as an additional variable in a set of Boltzmann equations
(even though it is not clear whether a thermally 
broadened resonance can be accurately
treated as an on-shell degree of freedom). 
Here we follow a different avenue, aiming eventually to offer for 
a framework permitting to scrutinize 
the accuracy of the Boltzmann approach.
For this purpose, we take the general linearization of 
\eq\nr{Boltzmann} as a starting point. 

%
\subsection{Basic definitions}

We next recall how $\Gamma_\rmi{chem}$ can be defined, 
through a linear response type analysis, as a transport 
coefficient~\cite{chemical}, and how its subsequent evaluation 
reveals the presence of a Sommerfeld effect~\cite{pert}. 
We only give the main steps, referring to refs.~\cite{chemical,pert}
for details. 

Let $\psi$ denote the Dirac spinor of a heavy quark.
Making use of a representation in which 
$\gamma^0 = \mbox{diag}(\unit^{ }_{2\times 2},-\unit^{ }_{2\times 2})$, 
$\psi$ can be expressed
in terms of spinors which have two non-zero components: 
\be
 \theta \equiv \fr12 (\unit + \gamma^0) \psi 
 \;, \quad
 \chi \equiv \fr12 (\unit - \gamma^0) \psi
 \;. \la{spinor}
\ee
In the following the vanishing components are omitted.
The spinor $\theta$ can be associated with a quark state, 
and the conjugate of the spinor $\chi$ with an antiquark state. 
In particular the energy density carried by heavy quarks and antiquarks, 
which for large $M$ is equal to their rest mass times their number density, 
can be described through 
\be
 H \; \equiv \; M(\theta^\dagger \theta - \chi^\dagger\chi)
 \; = \; M (\theta^*_{\alpha i} \theta^{ }_{\alpha i} + 
            \chi^{ }_{\alpha i} \chi^*_{\alpha i} )
 \;, \la{Hdef}
\ee
where $\theta^{ }_{\alpha i}$ annihilates a quark of colour $\alpha$
and spin $i$, $\chi^*_{\alpha i}$ does the same for an antiquark, 
and $M \equiv M^{ }_\rmi{rest}$ is the heavy quark rest mass.\footnote{%
 In perturbative considerations, particularly in \se\ref{se:pert}, 
 we make no difference between rest and kinetic masses, 
 assuming that both correspond to a pole mass. It is worth
 noting that the precise definition in \eq\nr{Hdef} is irrelevant since
 the mass cancels in \eq\nr{Gamma_def}; only the kinetic mass is 
 relevant for the Sommerfeld factors to be defined presently. 
 However the rest 
 mass strongly affects the quark number susceptibility $\chi^{ }_\ff$, 
 appearing e.g.\ in \eqs\nr{dG1} and \nr{dG8}, and 
 it also appears trivially in \eqs\nr{O1} and \nr{O8}. 
 } 
Note that we treat \eq\nr{Hdef}, without any terms suppressed
by $1/M$, as a {\em definition}; up to a normalization, it represents
a Noether charge density related to the NRQCD Lagrangian. 
 
We are concerned with the imaginary-time 
2-point correlator of the operator $H$, 
\ba
 \Delta(\tau) & \equiv &  \int_{\vc{x}} \,
 \Bigl\langle
   H(\tau,\vc{x}) \, H(0,\vc{0})
 \Bigr\rangle
 \;, \quad 0 < \tau < \frac{1}{T} 
 \;. \la{Delta_tau} 
\ea
Given that $H$ is a conserved charge density, 
$\Delta(\tau)$ is constant
within the standard NRQCD theory~\cite{nrqcd}. 
In full QCD, however, the 
number of heavy quarks is not conserved, because heavy quarks and
antiquarks 
can pair annihilate into gluons and light quarks. 
These processes can be described by adding 4-quark operators
to the NRQCD Lagrangian~\cite{bodwin}.

The 4-quark operators can be classified according to the gauge 
and spin quantum numbers of the 2-quark ``constituent operators'' 
of which they are composed (the 4-quark operators themselves
are gauge singlets and Lorentz  scalars). We carry out the 
main discussion in terms of the ``singlet'' operator, 
\be
 \mathcal{O}_1 ({}^{1}S^{}_0) \;\equiv\;
  \theta^\dagger\chi \, \chi^\dagger\theta
 \;, \quad 
 \delta S^{ }_\rmii{M} = 
 \frac{f_1 ({}^{1}S^{}_0)\, \mathcal{O}_1 ({}^{1}S^{}_0)}{M^2}
 \;, \la{O1}
\ee
whose coefficient has an ``absorptive'' 
imaginary part at $\rmO(\alphas^2)$: 
\be
 \im f_1 ({}^{1}S^{}_0) = \frac{\CF}{2\Nc} \pi \alphas^2 
 \;, \quad
 \CF \equiv \frac{\Nc^2 -1}{2\Nc}
 \;. 
\ee
For future
reference let us also write down one of the octet operators, 
\be
 \mathcal{O}_8 ({}^{1}S^{}_0) \equiv 
 \theta^\dagger T^a \chi \, \chi^\dagger T^a \theta
 \;, \quad 
 \delta S^{ }_\rmii{M} = 
 \frac{f_8 ({}^{1}S^{}_0)\, \mathcal{O}_8 ({}^{1}S^{}_0)}{M^2}
 \;, \la{O8}
\ee
where $T^a$ are generators of SU($\Nc$), normalized as 
$\tr(T^a T^b) = \delta^{ab}/2$. 
We work to leading non-trivial order in $\alphas(2 M)$ whereby
$\rmO(\alphas^3)$ corrections to the coefficients 
(cf.\ e.g.\ ref.~\cite{nlo3})
as well as operators suppressed by 
the relative velocity $v^2 \sim \alphas^2$
(cf.\ e.g.\ ref.~\cite{rel}) are omitted.  

If the singlet operator is added to the NRQCD Lagrangian, and the
correlator of \eq\nr{Delta_tau} is computed within this theory, then
$\Delta(\tau)$ is no longer a constant. We assume that the correlator  
is computed to {\em first order} as an expansion in 
$\im f_1 ({}^{1}S^{}_0) / M^2$.
If the result is Fourier decomposed, 
\be
 \tilde \Delta(\omega_n) = \int_0^\beta \! {\rm d}\tau \, 
 e^{i \omega_n \tau} \, \Delta(\tau) 
 \;, \quad 
 \beta \; \equiv \; \frac{1}{T} 
 \;, \la{Delta_wn}
\ee
and a corresponding spectral function is determined, 
$
 \rho^{ }_\Delta(\omega) = \im 
 \tilde \Delta(\omega_n \to -i [\omega+i0^+])
$, 
then a ``transport coefficient'' can be defined as
\be
 \Omega_\rmi{chem} \;\equiv\;
 \lim_{ \omega \ll T} 2 T \omega  \rho^{ }_\Delta(\omega)
 \;. \la{Omega_chem}
\ee
Note that the transport coefficient 
is extracted from a $1/\omega$ tail of the spectral function, 
rather than from a linear slope as is sometimes the case; 
the reason is that the interaction responsible for the process 
considered has been treated as an insertion.
The heavy quark chemical equilibration rate is subsequently 
obtained from~\cite{chemical} 
\be
 \Gamma_\rmi{chem} = \frac{\Omega_\rmi{chem}}{2 \chi^{ }_\ff M^2}
 \;, \la{Gamma_def}
\ee
where $\chi^{ }_\ff$ is the quark-number susceptibility related to 
the heavy flavour. Note that $M$ appears linearly in $H$
(cf.\ \eq\nr{Hdef}) and quadratically in $\Omega_\rmi{chem}$
(cf.\ \eq\nr{Delta_tau}) and therefore drops out in this ratio. 

The singlet operator in \eq\nr{O1} mediates decays which experience 
the so-called Sommerfeld enhancement~\cite{asommerfeld,landau3,fadin}. 
Within perturbation theory, omitting bound state effects, the
quark-number susceptibility $\chi^{ }_\ff$ and the singlet contribution 
to $\Omega_\rmi{chem}$, denoted by $\delta_1\Omega_\rmi{chem}$, 
read~\cite{pert} 
\ba
 \chi^{ }_\ff  & \approx & 4 \Nc \int_{\vc{p}}  e^{-\beta E_p }
 \; \approx \; 4\Nc 
 \Bigl( \frac{MT}{2\pi} \Bigr)^{\fr32} e^{-M/T} 
 \;, \la{pert1} \\ 
 \delta_1 \Omega_\rmi{chem} & \approx &  32 \Nc \im f_1 ({}^{1}S^{}_0)
 \int_{\vc{p,q}}
 e^{-\beta(E_p + E_q)}
 \, \bar{S}^{ }_\rmi{1} 
 \;,  \la{pert2}
\ea
where  $\bar{S}^{ }_\rmi{1}$ is a thermal Sommerfeld factor, related
to a vacuum Sommerfeld factor $S^{ }_1$ by 
\be
 \bar{S}^{ }_1 \simeq \frac{\int_{\vc{v}} e^{ -\beta E^{ }_\rmii{rel} } 
\,  S^{ }_1(v)}
 {\int_{\vc{v}} e^{ -\beta E^{ }_\rmii{rel}} }
 \;. \la{barS1_def}
\ee 
Here $E^{ }_\rmi{rel} \equiv M v^2 $, with $v = |\vc{v}|$, is the energy
of the relative motion of the annihilating 
particles. The normalization is such that without 
any resummation, $\bar{S}^{ }_\rmi{1} = S^{ }_1 = 1$. 

When higher-order perturbative corrections to $S^{ }_1$ are considered, 
there is particular subseries of them which proceeds in powers of $\alphas/v$.
These are summed to all orders into the Sommerfeld factor $S^{ }_1$. 
After the thermal average in \eq\nr{barS1_def}, $v$ is parametrically
of order $\sqrt{T/M}$. Therefore the thermal Sommerfeld effect 
is important for $T \lsim \alphas^2 M$, whereas it
evaluates to unity for $T \gg \alphas^2 M$.

Inspired by \eqs\nr{pert1} and \nr{pert2}, 
we may define a non-perturbative Sommerfeld factor as 
\be
 \bar{S}^{ }_\rmi{1} \, \equiv \,
 \frac{\Nc\, \delta_1 \Omega_\rmi{chem} }
  {2 \im f_1 ({}^{1}S^{}_0)\, \chi^{2}_\ff } 
 \;, \la{S1_nonpert}
\ee
where $\delta_1 \Omega_\rmi{chem}$ is to be evaluated with the operator
in \eq\nr{O1} inserted to first order. Note that the coefficient
$\im f_1 ({}^{1}S^{}_0)$ drops out in this ratio because
$\delta_1 \Omega_\rmi{chem}$ is linear in it. 
With this definition the chemical 
equilibration rate from \eq\nr{Gamma_def} can be expressed as 
\be
  \delta_1 \Gamma_\rmi{chem} =  \frac{ \im f_1 ({}^{1}S^{}_0) }{M^2} 
  \, \times \, \chi^{ }_\ff
  \, \times \, \frac{\bar{S}^{ }_\rmi{1}}{\Nc}
  \;. \la{dG1}
\ee
Thereby the physical result has factorized into 
a high-energy part $\im f_1 ({}^{1}S^{}_0)$ as well as ingredients
which can be addressed non-perturbatively. These are $\bar{S}^{ }_\rmi{1}$,
which can be extracted from a dimensionless ratio of 
correlation functions (cf.\ \eq\nr{barS1_final}), and the susceptibility
$\chi^{ }_\ff$, which is a standard observable measuring essentially
the equilibrium density $n_\rmi{eq}$. The appearance of $1/\Nc$
in \eq\nr{dG1} is a consequence of including $\Nc$ in $\chi^{ }_\ff$.

For the octet channel, the analogue of \eq\nr{pert2} reads~\cite{pert}
\be
 \delta_8 \Omega_\rmi{chem}  \; \approx \; 16 (\Nc^2-1) \im f_8 ({}^{1}S^{}_0)
 \int_{\vc{p,q}}
 e^{-\beta(E_p + E_q)}
 \, \bar{S}^{ }_\rmi{8} 
 \;, \la{pert2_S8}
\ee
where the perturbative value of $\bar{S}^{ }_8$ is given by an average 
like in \eq\nr{barS1_def}.  
We now define the non-perturbative value through 
\be
 \bar{S}^{ }_\rmi{8} \; \equiv \;
 \frac{\Nc^2\, \delta_8 \Omega_\rmi{chem} }
  {(\Nc^2 - 1) \im f_8 ({}^{1}S^{}_0)\, \chi^{2}_\ff } 
 \;. \la{S8_nonpert}
\ee
For the chemical equilibration rate this leads to 
\be
  \delta_8 \Gamma_\rmi{chem} =  \frac{ \im f_8 ({}^{1}S^{}_0) }{M^2} 
  \, \times \, \chi^{ }_\ff
  \, \times \, \frac{(\Nc^2 - 1)\bar{S}^{ }_\rmi{8}}{2 \Nc^2}
 \;. \la{dG8}
\ee
The colour factors follow from the definition of the octet operator
in \eq\nr{O8} and from the inclusion of one $\Nc$ in $\chi^{ }_\ff$.
In QCD there is also another octet operator
at $\rmO(\alphas^2, v^0)$, whose
coefficient $\im f_8 ({}^{3}S^{}_1)$ is proportional to 
the number of light flavours $\Nf$~\cite{bodwin};  for 
brevity we concentrate on spin-independent operators 
here but do not expect any qualitative
changes from spin-dependent terms. 

%
\section{Reduction to a static 2-point correlator}
\la{se:theory}

When the correlator in \eq\nr{Delta_tau} is evaluated by 
inserting the operator of \eq\nr{O1} to first order, we are 
faced with a 3-point function. We start by recalling that 
in the imaginary-time formalism the 3-point function 
splits up into a 4-point function. However, we subsequently
show that the 4-point function
can be reduced to a simple 2-point function, if we
are only interested in extracting the transport coefficient 
$\Omega_\rmi{chem}$ and not the full spectral shape. 

%
\subsection{Absorptive parts of 4-quark operators in imaginary time}
\la{se:imtime}

The imaginary parts of the 4-fermion vertices in NRQCD represent
pair annihilations of heavy quarks and antiquarks. Going to the 
center-of-mass frame, the vertex is a function of the total energy $E$ 
of the annihilating pair. Expanding the energy
dependence in powers of $E - 2M \approx M v^2 $, where 
$v$ denotes the velocity of $q$ 
in the $q\bar{q}$ rest frame, the leading term is a constant, 
i.e.\ of $\rmO(v^0)$. We restrict 
to the leading order in the non-relativistic expansion and thereby
only consider the constant term. 

Even though the term 
is constant, the corresponding vertex 
is non-local in imaginary time~\cite{chemical}. In general, the imaginary 
part corresponds to a spectral function, or cut, of a certain 
Green's function. More specifically, the 4-fermion vertex
can be viewed as a limit of a correlator of two quark-antiquark
operators. Given a spectral function for the latter, $\rho(E)$, the 
corresponding imaginary-time correlator is given by
\be
 G(\tau)
 = 
 \int_0^\infty \! \frac{{\rm d}E}{\pi} \, \rho(E) \, 
 \nB{}(E) \,
 \Bigl[e^{(\beta - |\tau|)E} + e^{|\tau|E} \Bigr]
 \;, \quad
 -\beta < \tau < \beta
 \;, \la{spec_rep_1}
\ee
where $\nB{}$ is the Bose distribution. Given that we know $\rho$
only for $E \approx 2 M$, we cut off the low-$E$ contribution by
a mass scale $\Lambda$, 
lying parametrically in the range $T \ll \Lambda \ll 2 M$. 
Denoting the constant value by $\rho_0$, 
we may then estimate
\be
 G(\tau) 
 \,\approx\,
 \int_{\Lambda}^\infty \! \frac{{\rm d}E}{\pi} \, \rho_0 \, 
 \nB{}(E) \,
 \Bigl[e^{(\beta - |\tau|)E} + e^{|\tau|E} \Bigr]
 \;\approx\; 
 \frac{\rho_0}{\pi} 
 \frac{e^{-|\tau|\Lambda}}{|\tau|} 
 \;\approx\; 
 \frac{\rho_0}{\pi |\tau|} 
 \;, \la{Gtau_2}
\ee
where we approximated 
$|\tau| \sim 1/(2M) \ll \beta$, $\exp({-\Lambda/T}) \ll 1$, 
and  $|\tau|\Lambda \sim \Lambda/(2M) \ll 1 $. 
Therefore, rather than obtaining 
$\delta(t)$ as would be the case in real time, 
we now get a non-locality $1/|\tau|$.
Specifically, the imaginary part of the 
singlet operator of \eq\nr{O1} can be expressed
as a {\em real} effective operator in a Euclidean action, 
\be
 S^{ }_\rmii{E} \approx - \frac{\im f_1({}^{1}S_0)}{\pi M^2}
 \int_0^\beta \! {\rm d}\tau_1  \,
 \int_0^\beta \! {\rm d}\tau_2  \,
 \int_{\vc{x}} 
 \, \frac{ (\theta^\dagger\chi) (\tau_1,\vc{x})
 \, (\chi^\dagger \theta)(\tau_2,\vc{x}) \, 
         }{|\tau_1 - \tau_2|}
 \;. \la{SE}
\ee

We remark that for the argumentation in \eq\nr{Gtau_2} it was 
essential that $1/|\tau|$ originated from the contribution
of large energies and corresponds therefore to a short separation. 
In \eq\nr{SE} this restriction has been lifted; this is not
a concern because, as our subsequent analysis shows, the contribution
still effectively emerges from short separations (cf.\ \eq\nr{master}).
A related point is that it is not clear how \eq\nr{SE} 
should be defined at the contact point $\tau_1 = \tau_2$.
Fortunately, the contact point is naturally
regularized by the considerations to follow, so  for the moment
we simply treat \eq\nr{SE} as a formal construction. 

%
\subsection{Canonical analysis and analytic continuation}

Expanding now $\exp({- S^{ }_\rmii{E}}) \to 1 - S^{ }_\rmii{E}$,  
defining
\be
 \Delta(\tau) \, \equiv \, \frac{\im f_1({}^{1}S_0)}{\pi M^2}\, \Epsilon(\tau) 
 \;, \la{Epsilon_def}
\ee
and making use of translational invariance in order to adjust the spatial
coordinates, \eqs\nr{Delta_tau} 
and \nr{SE} imply that we are faced with the 4-point correlator 
\be
 \Epsilon(\tau) 
 \; \equiv \;  \int_{\vc{x},\vc{y}}
   \int_{0}^\beta\! {\rm d}\tau_2 \, \int_{0}^{\beta} \! {\rm d}\tau_1 \, 
   \frac{\langle H(\tau,\vc{x})\, H(0,\vc{y})\,
   (\theta^\dagger\chi) (\tau_1,\vc{0})
   \, (\chi^\dagger \theta)(\tau_2,\vc{0})
   \rangle}{|\tau_1 - \tau_2|}
 \;. \la{Eps_tau}
\ee
We proceed to giving a canonical interpretation to \eq\nr{Eps_tau}, 
showing that the 4-point correlator
reduces to a 2-point correlator once
we extract the corresponding transport coefficient. 

We reiterate, first of all, that within the NRQCD Lagrangian $\int_\vc{x} H$
corresponds to a conserved charge 
(particle plus antiparticle number times the rest mass). 
In other words, the corresponding operator
commutes with the Hamiltonian $\hat{\mathcal{H}}$, 
$ [\hat{\mathcal{H}}, \int_\vc{x} \hat{H} ] = 0 $. 
Therefore the Heisenberg operator
is time independent, 
\be
 \int_\vc{x} \hat{H}(\tau,\vc{x}) = 
 \int_\vc{x} \hat{H}(\tau',\vc{x})  
 \;. 
 \la{time-indep}
\ee
It turns out, however, that there are contact terms at the positions
$\tau_1$ and $\tau_2$ when considering the correlator in \eq\nr{Eps_tau}. 
For $\tau_1 > \tau_2$, which will turn
out to be the case relevant for us (see below), 
we need to have $\tau > \tau_1$ or
$\tau < \tau_2$, otherwise the correlator vanishes up to exponentially
small corrections.  Therefore $\tau$ can only be chosen from 
the range $\tau \in [0,\tau_2) \cup (\tau_1,\beta] $.

In order to re-express \eq\nr{Eps_tau} in the canonical formalism
in this case, 
we recall that the Euclidean path integral corresponds to a time-ordered
expectation value. 
Now, within the NRQCD theory, the operators 
$\hat{\theta}^\dagger$ and $\hat{\theta}$ are the creation and annihilation 
operators for the quark states, respectively. For the antiquark 
states, $\hat{\chi}$ corresponds to a creation operator and 
$\hat{\chi}^\dagger$
to an annihilation operator. Therefore $\hat{\theta}^\dagger\hat{\chi}$ 
creates a quark-antiquark state, and $\hat{\chi}^\dagger\hat{\theta}$ 
annihilates it. According to  \eq\nr{Hdef}, $\int_\vc{x} \hat{H}$
measures the rest mass of the quarks and antiquarks present. 

We denote energy eigenstates involving a heavy quark and a heavy
antiquark by $| q\bar{q},m \rangle$, whereas states involving neither
are denoted by $|n\rangle$. The corresponding energies are denoted 
by $E_m$ and $\epsilon_n$, respectively. Then 
$
 \int_\vc{x}  \hat H \, | q\bar{q},m \rangle = 
 2 M | q\bar{q},m \rangle
$ 
whereas 
$
 \int_\vc{x}  \hat H \, | n \rangle = 0
$.
Note that 
the quark-antiquark states can be either bound states or scattering states. 

Inserting now completeness relations\footnote{%
 For notational simplicity we assume that the system is placed for a moment
 in a finite volume so that the energy spectrum is discrete. 
 } 
and making use of \eq\nr{time-indep}, 
the numerator of \eq\nr{Eps_tau} can be expressed as 
\ba
 && \hspace*{-1cm}
 \int_{\vc{x},\vc{y}}
 \left. \langle H(\tau,\vc{x})\, H(0,\vc{y})\,
   (\theta^\dagger\chi) (\tau_1,\vc{0})
 \, (\chi^\dagger \theta)(\tau_2,\vc{0})
 \rangle \right|^{ }_{\tau_1 > \tau_2} 
 \bigl[ \theta(\tau - \tau_1) + \theta(\tau_2 - \tau) \bigr]
 \nn 
 & = & 
 \frac{1}{\mathcal{Z}} \,
 \tr \Bigl[e^{-\beta \hat{\mathcal{H}}}
  (\hat\theta^\dagger\hat\chi)(\tau_1,\vc{0}) \, 
  (\hat\chi^\dagger\hat\theta)(\tau_2,\vc{0}) \, 
 \int_{\vc{x},\vc{y}} \hat{H}(0,\vc{x}) \hat{H}(0,\vc{y})
  \Bigr]
 \nn 
 & = & 
 \frac{1}{\mathcal{Z}} \, \sum_{m,n}
 \langle q\bar{q},m | e^{-\beta \hat{\mathcal{H}}}
  (\hat\theta^\dagger\hat\chi)(\tau_1,\vc{0}) \, |n\rangle\,\langle n |  
  (\hat\chi^\dagger\hat\theta)(\tau_2,\vc{0}) \, 
 \int_{\vc{x},\vc{y}} \hat{H}(0,\vc{x}) \hat{H}(0,\vc{y})
 | q\bar{q},m \rangle
 \nn 
 & = & 
 \frac{ 4 M^2 }{\mathcal{Z}} 
 \sum_{m,n} e^{-\beta E_m} e^{(\tau_1 - \tau_2)(E_m - \epsilon_n )}
 \, \langle q\bar{q},m | \hat\theta^\dagger\hat\chi | n \rangle 
     \langle n | \hat\chi^\dagger\hat\theta | q\bar{q},m \rangle 
 \;. \la{canonical}
\ea
We have restricted to the contribution from sectors of the Hilbert space
with at most one quark-antiquark state present. The quark-antiquark states
had to be placed at the outer boundaries in order for $\int_\vc{x}\hat H$
to give a non-zero contribution. 

We note that had we chosen $\tau_2 > \tau_1 $ instead, 
then the creation operator 
$
 (\hat\theta^\dagger\hat\chi)(\tau_1,\vc{0})
$
would have 
appeared to the right of  the annihilation operator
$
 (\hat\chi^\dagger\hat\theta)(\tau_2,\vc{0})
$.
Then the states $|n\rangle$ would have to 
be replaced by states containing two heavy quarks and two antiquarks. 
This is the case irrespective of the range chosen for $\tau$, because
one of the $\hat{H}$-operators is always at $\tau = 0$.
The corresponding contribution to $\Gamma^{ }_\rmi{chem}$ would be 
exponentially suppressed by $\exp(-2M/T)$ and will be omitted. 

Eq.~\nr{canonical} should clarify the physical meaning 
of \eq\nr{Eps_tau}. As indicated by the matrix elements squared, 
we are considering the decays of quark-antiquark
states into light degrees of freedom. Physically, 
the quark-antiquark states are 
assumed to be initially {\em close to} thermal equilibrium; 
our general framework corresponds to a linear response analysis, 
so that in \eq\nr{canonical} the quark-antiquark 
states are {\em in} thermal equilibrium. 

Understanding the time dependence of \eq\nr{canonical} 
requires care. Let us denote
\be
 \mathcal{C}^{ }_{mn} \; \equiv \;  
 \frac{ 4 M^2 }{\mathcal{Z}} 
 e^{-\beta E_m} 
 \, \langle q\bar{q},m | \hat\theta^\dagger\hat\chi | n \rangle 
     \langle n | \hat\chi^\dagger\hat\theta | q\bar{q},m \rangle 
 \;. \la{Cmn}
\ee
Recalling the restrictions on $\tau$
the function in \eq\nr{Eps_tau} can then be written as 
\be
 \mathcal{E}(\tau) =
 \sum_{m,n}  \mathcal{C}^{ }_{mn}  \int_{0}^\beta\! {\rm d}\tau_2 \, 
  \int_{\tau_2}^{\beta} \! {\rm d}\tau_1 \,  
  \frac{e^{(\tau_1 - \tau_2)(E_m - \epsilon_n)}}{\tau_1 - \tau_2} \,
  \bigl[  \theta(\tau - \tau_1) + \theta(\tau_2 - \tau) \bigr]
  \;. 
\ee
Taking a time derivative; carrying out one integration with the help
of the resulting Dirac $\delta$-function; and representing  
the denominator as 
$  
  1/(\tau_1 - \tau_2) =
  \int_0^\infty \! {\rm d}s \, e^{-s (\tau_1 - \tau_2)}
$, 
we get
\ba
 \mathcal{E}'(\tau) & = & -\,  \sum_{m,n}  \mathcal{C}^{ }_{mn}
 \int_{\tau}^{\beta-\tau} \! 
 \frac{{\rm d}x}{x}\, e^{x(E_m - \epsilon_n)} 
 \nn 
 & = & -\,  \sum_{m,n}  \mathcal{C}^{ }_{mn} 
 \int_{\tau}^{\beta-\tau} \! {\rm d}x \, 
 \int_0^\infty \! {\rm d}s \, e^{x(E_m - \epsilon_n - s)} 
 \nn 
 & = & 
  \sum_{m,n}  \mathcal{C}^{ }_{mn}
 \int_0^\infty \! {\rm d}s \,
 \frac{e^{\tau(E_m - \epsilon_n - s)}
   - e^{(\beta - \tau)(E_m - \epsilon_n - s)}
  }{E_m - \epsilon_n - s}
 \;.
\ea
This can be integrated into
\be
 \mathcal{E}(\tau) =  \sum_{m,n}  \mathcal{C}^{ }_{mn} 
 \int_0^\infty \! {\rm d}s \,
 \frac{e^{\tau(E_m - \epsilon_n - s)}
   + e^{(\beta - \tau)(E_m - \epsilon_n - s)}
  }{(E_m - \epsilon_n - s)^2}
  - \mbox{const.}
 \;, 
\ee
where the (infinite) integration constant can be omitted, because
its Fourier transform has no non-trivial cut. 
Going over to the normalization of \eq\nr{Epsilon_def}, 
a Fourier transform according to \eq\nr{Delta_wn} yields 
\be
 \tilde\Delta(\omega_n) = 
 \frac{\im f_1({}^{1}S_0)}{\pi M^2}\, \sum_{m,n}  \mathcal{C}^{ }_{mn}
 \, 
 \int_0^\infty \! {\rm d}s \, 
 \frac{
  e^{\beta(E_m - \epsilon_n - s)} - 1
 }{(E_m - \epsilon_n - s)^2}
 \, \biggl[ \frac{1}{i\omega_n + E_m - \epsilon_n - s } +
  (\omega_n \to - \omega_n )\biggr]
 \;. 
\ee
The corresponding spectral function reads
\ba
 \rho^{ }_{\Delta}(\omega) \hspace*{-6mm} & = & \hspace*{-6mm} 
 \frac{\im f_1({}^{1}S_0)}{M^2}\, \sum_{m,n}  \mathcal{C}^{ }_{mn}
 \, 
 \int_0^\infty \! {\rm d}s \, 
 \frac{
  e^{\beta(E_m - \epsilon_n - s)} - 1
 }{(E_m - \epsilon_n - s)^2}
 \,\bigl[\delta(-\omega + E_m - \epsilon_n - s) - (\omega\to -\omega) \bigr]
 \nn 
 & \stackrel{\omega,\epsilon_n \ll E_m }{=} & \hspace*{-2mm}
 \frac{\im f_1({}^{1}S_0)}{M^2}\, \sum_{m,n}  \mathcal{C}^{ }_{mn}
 \, 
 \frac{e^{\beta\omega} - e^{-\beta\omega}}{\omega^2}
 \;,  
\ea
where we noted that the dynamical energy scales contained in the  
effective description, reflected by the energy eigenvalues 
$\epsilon_n$, are by assumption much smaller
than the heavy quark-antiquark energies $E_m \sim 2 M$, so 
that the contribution emerges from $s\sim 2 M$.\footnote{%
 This is a subtle point. It could be said that the energies $E_m \sim 2 M$
 had already been integrated out in order 
 to arrive at the NRQCD description. However,
 in order to get the correct limit, we need to keep the rest mass in the
 heavy quark Lagrangian here (cf.\ \eq\nr{propG}); 
 it can only be shown {\it a posteriori} (see below)
 that the rest mass can be shifted away also in our 
 finite-temperature observables. 
 }
Taking finally the limit in \eq\nr{Omega_chem} and inserting
subsequently the value of $\mathcal{C}^{ }_{mn}$ from \eq\nr{Cmn} yields 
\ba
 \Omega_\rmi{chem} & = & 
 \frac{4 \im f_1({}^{1}S_0)}{M^2}\, \sum_{m,n}  \mathcal{C}^{ }_{mn}
 \nn[2mm] 
 & = & 
 16 \,
 \im f_1({}^{1}S_0)\,
 \frac{1}{\mathcal{Z}} 
 \sum_{m,n} e^{-\beta E_m} 
 \, \langle q\bar{q},m | \hat\theta^\dagger\hat\chi | n \rangle 
     \langle n | \hat\chi^\dagger\hat\theta | q\bar{q},m \rangle 
 \nn 
 & = & 
 16 \,
 \im f_1({}^{1}S_0)\, \frac{1}{\mathcal{Z}}
 \, \tr \Bigl[ e^{-\beta\hat{\mathcal{H}} }
  (\hat\theta^\dagger\hat\chi)(0^+,\vc{0}) \, 
  (\hat\chi^\dagger\hat\theta)(0,\vc{0})
 \Bigr]
  \nn[2mm] 
 & = & 
 16 \,
 \im f_1({}^{1}S_0)\,
 \bigl\langle 
  (\theta^\dagger\chi)(0^+,\vc{0}) \, 
  (\chi^\dagger\theta)(0,\vc{0})
 \bigr\rangle 
 \;. \la{master}
\ea
Here an essential point was that there was no
functional dependence on the energy eigenvalues~$\epsilon_n$, 
so that we could re-identify
the sum $\sum_n |n\rangle\langle n |$ as a unit operator.
Subsequently we returned to a path integral representation.  
We also indicated one of the time arguments by $0^+$ in order to maintain
the correct ordering. 

Eq.~\nr{master} is one of our main results. 
Effectively, it represents the 
expectation value of the imaginary part of 
the singlet operator in \eq\nr{O1}, 
albeit in a spacetime
with a Euclidean signature, with a periodic time direction, 
and with a specific time ordering.
Eq.~\nr{master} shows that even though 
the general representation of the absorptive parts of 4-quark operators 
is cumbersome (cf.\ \se\ref{se:imtime}), 
the final transport coefficient {\em can} be extracted from an ``almost''
local 4-fermion operator. 
The time ordering in \eq\nr{master}, represented by the time argument $0^+$,  
has however a specific consequence, to which we now turn.

%
\subsection{Wick contractions}
\la{ss:wick}

In order to express \eq\nr{master} in terms of propagators, 
we need to recall some of their basic properties. The heavy quark 
propagator is defined in appendix~A. An important consequence of the fact
that non-relativistic propagators are defined by equations which
are of first order in time is that 
$G^\theta_{ }(\tau_2,\vc{x};\tau_1,\vc{y}) = 
\langle \theta(\tau_2,\vc{x}) \theta^\dagger(\tau_1,\vc{y})\rangle $ 
is obtained
by integrating into the region $\tau_2 > \tau_1$ (cf.\ \eq\nr{propG}), 
and $G^\chi_{ }(\tau_2,\vc{x};\tau_1,\vc{y}) = 
\langle \chi(\tau_2,\vc{x}) \chi^\dagger(\tau_1,\vc{y})\rangle $ is obtained
by integrating into the region $\tau_1 > \tau_2$.
However in \eq\nr{master} the time argument of 
$\theta^\dagger$ is larger than that of
$\theta$, and the time argument of 
$\chi$ is larger than that of $\chi^\dagger$. 
Therefore a non-zero contraction requires integrating around
the imaginary-time direction. 
The situation is illustrated in \fig\ref{fig:Gcirc}. 

%
\begin{figure}[t]
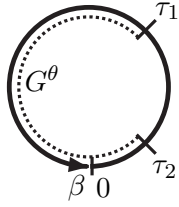


\hspace*{2.2cm}%
\begin{minipage}[c]{11.2cm}
\begin{eqnarray*}
&& 
 \hspace*{-1cm}
 \Gcirc  
\end{eqnarray*}
\end{minipage}

\vspace*{2mm}

\caption[a]{\small 
 The propagator 
 $G^\theta(\tau_2,\vc{0};\tau_1,\vc{0})$ for the 
 ``unnatural'' time ordering $\tau_1 > \tau_2$.  
 In this case the propagator represents a movement 
 across the point $\tau = \beta$.
 The circle stands for the imaginary time direction. 
 At zero temperature ($\beta\to\infty$) the propagator 
 would be exponentially suppressed. 
 This structure corresponds to \eq\nr{canonical}, 
 where only light parton states appear between $\tau_2$ and $\tau_1$.
 }
\la{fig:Gcirc}
\end{figure}
%

With these observations and the relations
in \eqs\nr{bcG} and \nr{Gchi} in mind, 
the correlator in \eq\nr{master}
can be expressed as
\ba
 \bigl\langle 
  (\theta^\dagger\chi)(0^+,\vc{0}) \, 
  (\chi^\dagger\theta)(0,\vc{0})
 \bigr\rangle 
 & = & 
 - \Bigl\langle G^\theta_{\alpha\gamma;ij}(0,\vc{0};0^+,\vc{0}) \,
   G^\chi_{\gamma\alpha;ji}(0^+,\vc{0};0,\vc{0}) \Bigr\rangle
 \nn
 & = & 
 - \Bigl\langle G^\theta_{\alpha\gamma;ij}(\beta,\vc{0};0^+,\vc{0}) \,
   G^\chi_{\gamma\alpha;ji}(0^+,\vc{0};\beta,\vc{0}) \Bigr\rangle
 \nn
 & = & \tr \Bigl\langle 
   G^\theta(\beta,\vc{0};0,\vc{0}) \,
   G^{\theta\dagger}(\beta,\vc{0};0,\vc{0}) 
  \Bigr\rangle
 \;, \la{wick}
\ea
where $\alpha, \gamma \in\{1,...,\Nc\}$ are
colour indices, and $i,j \in\{1,2\}$ are spin indices.
We may also express $\chi^{ }_\ff$, entering 
\eq\nr{S1_nonpert}, in the same notation. A few manipulations, 
including the use of \eq\nr{Gchi}, lead to 
\ba
 \chi^{ }_\ff & = & 
 \int_\vc{x} 
 \Bigl\langle
  \bigl(\theta^\dagger\theta + \chi^\dagger\chi \bigr)(\tau,\vc{x})
  \bigl(\theta^\dagger\theta + \chi^\dagger\chi \bigr)(0,\vc{0})
 \Bigr\rangle
 \nn 
 & = & 
 - \int_\vc{x} \tr \Bigl\langle 
   G^\theta(0,\vc{0};\tau,\vc{x}) G^\theta(\tau,\vc{x};0,\vc{0}) + 
   G^\chi(0,\vc{0};\tau,\vc{x}) G^\chi(\tau,\vc{x};0,\vc{0}) 
 \Bigr\rangle
 \nn 
 & = & 
  \int_\vc{x} \tr \Bigl\langle 
   G^\theta(\beta,\vc{0};\tau,\vc{x}) G^\theta(\tau,\vc{x};0,\vc{0}) + 
   G^\chi(0,\vc{0};\tau,\vc{x}) G^\chi(\tau,\vc{x};\beta,\vc{0}) 
 \Bigr\rangle
 \nn 
 & = & 
 \tr\Bigl\langle
   G^\theta(\beta,\vc{0};0,\vc{0}) + 
   G^{\theta\dagger}(\beta,\vc{0};0,\vc{0}) 
 \Bigr\rangle
 \; = \; 
 2 \re \tr \bigl\langle 
   G^\theta(\beta,\vc{0};0,\vc{0}) \bigr\rangle 
 \;. \la{chiff_final}
\ea

Inserting \eq\nr{wick} into \eq\nr{master} and 
normalizing the result according to \eq\nr{S1_nonpert}, 
with $\chi^{ }_{\ff}$ inserted from \nr{chiff_final}, 
we get our final result for the non-perturbative 
Sommerfeld factor: 
\be
 \bar{S}_1 \; = \; 
 \frac{ \frac{1}{2 \Nc} \tr 
 \bigl\langle 
   G^{\theta}_{ } (\beta,\vc{0};0,\vc{0})
   G^{\theta\dagger}_{ } (\beta,\vc{0};0,\vc{0})  
 \bigr\rangle }
 {\bigl\{ \frac{1}{2 \Nc} \re \tr \bigl\langle 
   G^\theta(\beta,\vc{0};0,\vc{0})  \bigr\rangle \bigr\}^2}
 \;. \la{barS1_final} 
\ee
The factor $2\Nc$ corresponds to the dimension of the propagator matrices. 

The following physical interpretation can be suggested. Effectively we
are computing the thermal expectation value of the imaginary part of 
the 4-quark operator, cf.\ \eq\nr{master}. But the contractions are
non-trivial in that a quark-antiquark pair
is generated at time $0$ and annihilated at time $\beta$. The 
circling of the imaginary-time direction guarantees that the
corresponding physical states are thermalized.  
The system is allowed to decide whether the propagation
takes place in the form of open or bound 
states, as long as they are distributed
according to the proper thermal weights. 
This physics is normalized by the propagators of two 
independent heavy quarks or antiquarks. 

In order to write down the corresponding 
correlator for the octet case, \eq\nr{O8}, we define the ratios
\ba
 P^{ }_1 & \equiv & \frac{1}{2 \Nc} \re\, \bigl\langle 
   G^\theta_{\alpha\alpha;ii}(\beta,\vc{0};0,\vc{0})  \bigr\rangle 
 \;, \la{defP1} \\ 
 P^{ }_2 & \equiv & 
 \frac{1}{2\Nc} 
 \bigl\langle 
   G^{\theta}_{\alpha\gamma;ij } (\beta,\vc{0};0,\vc{0})
   G^{\theta\dagger}_{\gamma\alpha;ji} (\beta,\vc{0};0,\vc{0})  
 \bigr\rangle
 \;, \la{defP2} \\
 P^{ }_3 & \equiv & 
 \frac{1}{2\Nc^2} 
 \bigl\langle 
   G^{\theta}_{\alpha\alpha;ij} (\beta,\vc{0};0,\vc{0})
   G^{\theta\dagger}_{\gamma\gamma;ji } (\beta,\vc{0};0,\vc{0})  
 \bigr\rangle
 \;. \la{defP3}
\ea
Summing over the generators $T^a$ and normalizing 
the result according to \eq\nr{S8_nonpert} we get 
\be
 \bar{S}_8 \; = \; 
 \frac{ \Nc^2 P^{ }_3 - P^{ }_2 }
 {(\Nc^2 - 1) P_1^2 }
 \;. \la{barS8_final} 
\ee
With the same notation, $\bar{S}_1 = P^{ }_2 / P_1^2$.
Note that all the quantities entering the singlet and octet factors are 
manifestly gauge-invariant. 


%
\section{Perturbative estimates}
\la{se:pert}

In order to evaluate \eq\nr{barS1_final} within (resummed) 
perturbation theory, 
we rewrite it in the form of a spectral representation. 
Denoting by $\rho^{ }_{q\bar{q}}(E,k)$ the spectral density
of quark-antiquark states, and representing the spectral density
of the single-particle states in the denominator by the non-relativistic
free form $\rho^{ }_{q}(E,p) = 2\Nc \pi \delta(E- E^{ }_p)$, 
the spectral representation of \eq\nr{barS1_final} reads\footnote{%
 The numerator is analogous to the relativistic form in \eq\nr{spec_rep_1}; 
 the  non-relativistic contribution originates from the first term. 
 The peculiar nature of the Wick contractions discussed in \se\ref{ss:wick}, 
 related to the non-relativistic propagators used, 
 subsequently sets $\tau\to\beta$.
 } 
\be 
 \bar{S}^{ }_1 \; \approx  \;
 \frac{ 
 \frac{1}{2\Nc} \int_{\Lambda}^{\infty}\! \frac{{\rm d}E}{\pi}
 e^{-\beta E} \int_\vc{k} \rho^{ }_{q\bar{q}}(E,k) }{
 \int_{\vc{p},\vc{q}} e^{-\beta (E^{ }_p + E^{ }_q)}}
 \;. \la{spec_rep_2}
\ee
Here 
$\vc{k}$ is 
the momentum of the quark-antiquark pair with respect
to the heat bath. The Boltzmann weight $e^{-\beta E}$ implies
that states with a small mass appear with a high likelihood. 

\begin{figure}[t]

\hspace*{-0.1cm}
\centerline{%
 \epsfxsize=8.5cm\epsfbox{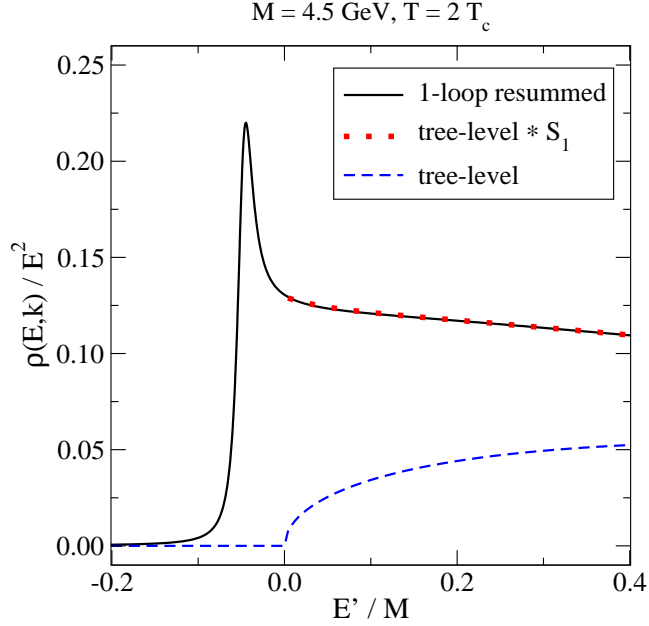}%
}

\caption[a]{\small
 The tree-level (cf.\ \eq\nr{tree_res}) and 1-loop resummed
 (cf.\ refs.~\cite{resum,peskin}) pseudoscalar spectral 
 function in hot QCD, with $E' \equiv E -2M - k^2 / (4 M)$
 denoting the energy with respect to the 2-particle threshold. 
 Here $k = |\vc{k}|$ is the spatial momentum with 
 respect to the heat bath.  The tree-level result multiplied
 by the Sommerfeld factor $S_1$ from \eq\nr{Ss} 
 reproduces the above-threshold spectral shape,  
 but it misses the (thermally broadened) resonance-like contribution 
 below the threshold. 
}

\la{fig:spectral}
\end{figure}

We evaluate \eq\nr{spec_rep_2} in three ways, with 
increasing sophistication. At tree-level, the spectral function originates
from free scattering states, and reads
\be
 \rho^{ }_{q\bar{q}}(E,k) \; \to \; \rho^{ }_\rmi{tree}(E,k) 
 \; \equiv \; 
 2\Nc \int_{\vc{p},\vc{q}} \pi \delta(E- E^{ }_p - E^{ }_q)
 \,(2\pi)^3 \delta^{(3)}(\vc{k-p-q})
 \;. \la{rho_tree} 
\ee
Inserting this into \eq\nr{spec_rep_2} and carrying out the integrals
over $E$ and $\vc{k}$, immediately yields $\bar{S}^{ }_1 = 1$.  
For future reference, it is helpful to express the tree-level result
in another way as well. Making use of the non-relativistic form 
$E^{ }_p = M + p^2 / (2 M)$, the integrals over $\vc{p,q}$
in \eq\nr{rho_tree} are readily carried out, yielding
\be
 \rho^{ }_\rmi{tree}(E,k) \; = \; 
 \frac{\Nc M^{\fr32} \theta(E') \sqrt{E'} }{2\pi}
 \;, \quad
 E' \; \equiv \; 
 E - 2 M - \frac{k^2}{4M}
 \;. \la{tree_res}
\ee
This result is illustrated with a dashed line in \fig\ref{fig:spectral}.
We note that it amounts to $-1/3$ times the vector channel 
spectral function studied
in ref.~\cite{resum}, i.e.\ the pseudoscalar channel~\cite{peskin}.

As a second step, we include the Sommerfeld enhancement in its standard
form. Employing 
QCD language, we refer to this approximation as a contribution
from ``open'' (or ``above-threshold'', or ``scattering'') states. Then 
\be
 \rho^{ }_{q\bar{q}}(E,k) \; \to \; \rho^{ }_\rmi{open}(E,k) 
 \; \equiv \; 
 2\Nc \int_{\vc{p},\vc{q}} \pi \delta(E- E^{ }_p - E^{ }_q)
 \,(2\pi)^3 \delta^{(3)}(\vc{k-p-q})\, S^{ }_1(v)
 \;, \la{rho_open} 
\ee
where $v = |\vc{v}|$ is the relative velocity defined through 
\be
 \vc{v} \; \equiv \; \frac{\vc{p-q}}{2M}
 \;. \la{v_def}
\ee
The Sommerfeld factor reads~\cite{fadin}
\be
  \Ss = \frac{ \Xs } { 1 - e ^{ - \Xs } }
  \;, \quad 
  \Xs   =  \frac{ g ^ 2 \CF  } { 4 v }
  \;. \label{Ss}
\ee
\Eqs\nr{rho_open} and \nr{v_def} imply that 
$E^{ }_p + E^{ }_q = 2 M + k^2/(4M) + M v^2$, so we can 
identify the variable $E'$ of \eq\nr{tree_res}
as $E' = M v^2$. Therefore $\rho^{ }_\rmi{open}$ can 
be evaluated by multiplying $\rho^{ }_\rmi{tree}$ from \eq\nr{tree_res}
by $S_1^{ }(v)$; the result is shown with a dotted line
in \fig\ref{fig:spectral}.

We note in passing that expressing 
both the numerator and denominator of \eq\nr{spec_rep_2}
in center-of-mass variables, and inserting $\rho^{ }_\rmi{open}$
from \eq\nr{rho_open}, a simple computation yields
\be
 \bar{S}^{ }_{1,\rmi{open}}
 = \frac{\int_\vc{v} e^{-\frac{M v^2}{T}} S^{ }_1(v) }
 { \int_\vc{v} e^{-\frac{M v^2}{T}} }
 \;. \la{barS1_2}
\ee
This agrees with the expression given in \eq\nr{barS1_def}. 

\begin{figure}[t]

\hspace*{-0.1cm}
\centerline{%
 \epsfxsize=7.5cm\epsfbox{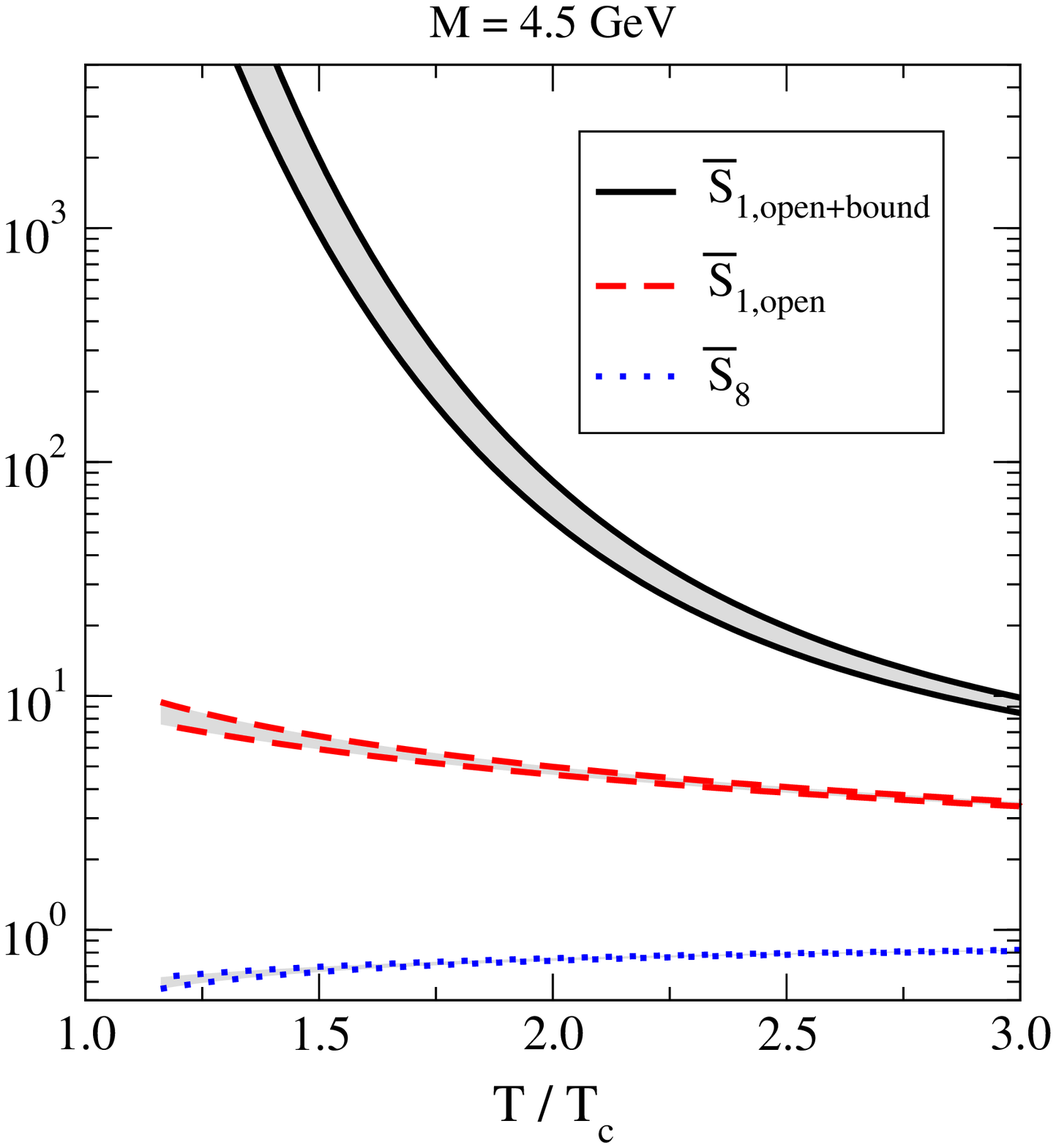}%
 \hspace{0.1cm}
 \epsfxsize=7.5cm\epsfbox{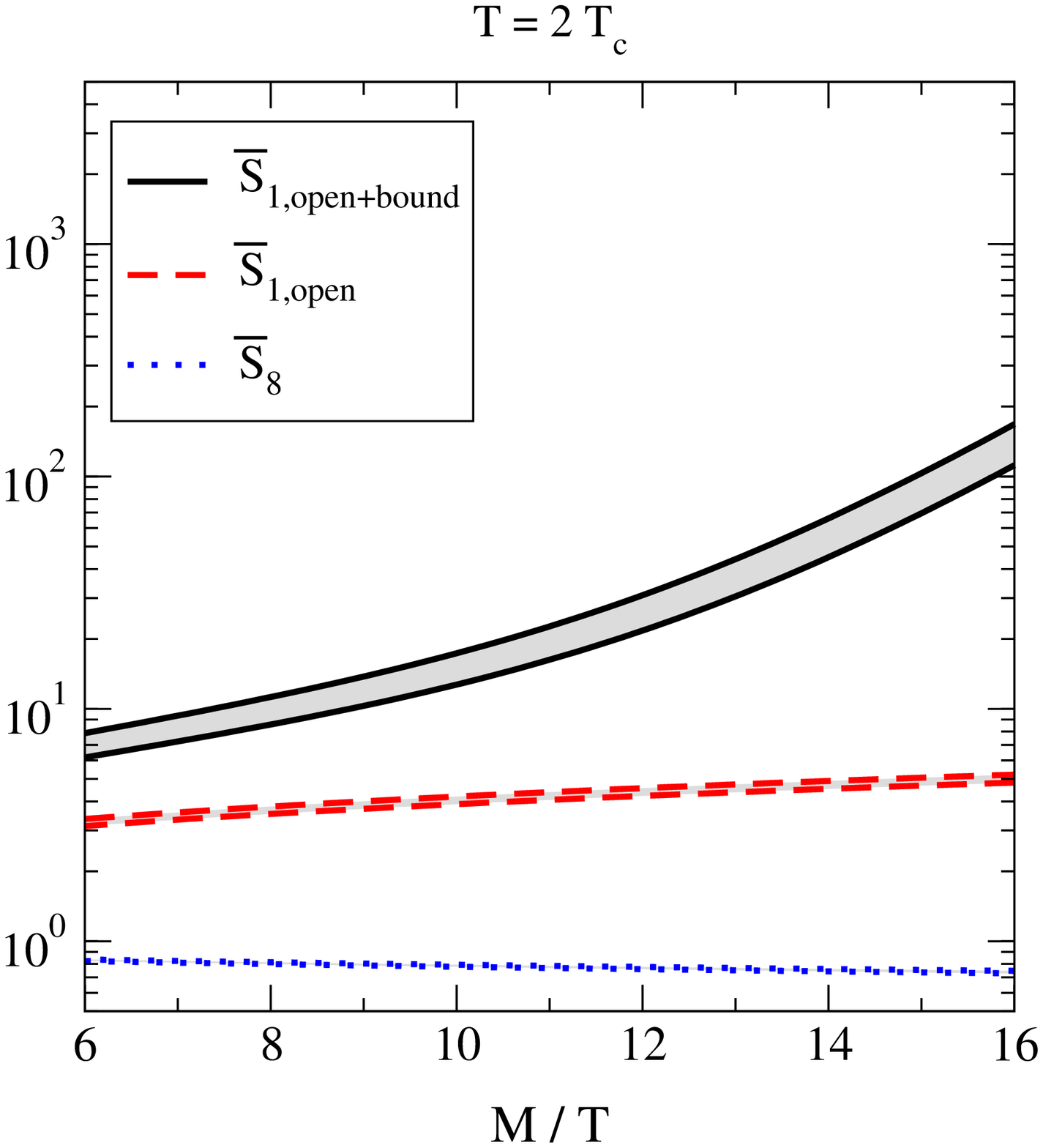}
}

\caption[a]{\small
 Perturbative Sommerfeld factors, as a function of $T$
 for fixed $M$ (left), and as a function of $M$
 for fixed $T$ (right). As a gauge coupling we 
 have used the 2-loop dimensionally reduced coupling 
 from ref.~\cite{gE2}, with 
 $\Lambdamsbar \simeq 360$~MeV~\cite{pacs}, 
 and moreover we chose $\Tc \simeq 155$~MeV 
 (for a review, see ref.~\cite{hbm}). 
 The numerical values vary in the range $\alphas \simeq 0.2...0.4$.
 The bands follow from variations of the renormalization scale 
 by a factor $1/2 ... 2$ around a typical thermal choice $\sim 2\pi T$, 
 but represent only a lower bound of theoretical uncertainties. 
}

\la{fig:pert}
\end{figure}

As a third level of sophistication, 
we evaluate \eq\nr{spec_rep_2} by including
the full spectral shape shown in \fig\ref{fig:spectral}. We refer
to this as a contribution from ``open+bound'' states. Substituting
variables as $E' = E - 2M - k^2 / (4M)$; noting that the 1-loop
resummed spectral function\footnote{%
 The 1-loop resummed spectral function corresponds to the imaginary
 part of a Coulomb Green's function, with the potential appearing in 
 the inhomogoneous Schr\"odinger equation computed up to order $g^2$ 
 in Hard Thermal Loop resummed perturbation theory. This potential 
 has a real part, accounting for 
 ``virtual'' thermal corrections such as Debye screening, 
 as well as an imaginary part, 
 accounting for  ``real'' 
 thermal corrections such as elastic $2\to 2$ scatterings with 
 light plasma constituents, which lead to thermal broadening.
 } 
depends on $E'$ only~\cite{resum},
so that we can write
$
 \rho^{ }_{q\bar{q}}(E,k) = {\rho}(E')
$; 
and carrying out both integrals in the denominator of \eq\nr{spec_rep_2}
as well as the integral over $\vc{k}$ in the numerator, yields
\be
 \bar{S}^{ }_{1,\rmi{open+bound}} 
 \; = \; 
 \frac{1}{2\Nc}
 \biggl(\frac{4\pi}{MT} \biggr)^{\fr32} 
 \int_{-\Lambda'}^{\infty}
 \! \frac{{\rm d}E'}{\pi} \, 
 e^{-\frac{E'}{T} } {\rho}(E')
 \;. \la{rho_open_bound}
\ee
The cutoff scale $\Lambda'$ has no practical significance, given that
the spectral function vanishes for $E' \ll - \alphas^2 M$
(cf.\ \fig\ref{fig:spectral}). 

A numerical evaluation of \eq\nr{rho_open_bound}, based on 
the spectral function shown in \fig\ref{fig:spectral}, 
is illustrated in \fig\ref{fig:pert}.
It is obvious that at low temperatures, the bound
state contribution completely dominates the result, 
boosting $\bar{S}^{ }_1$ by more than two orders of magnitude. 

In the octet case, where the interaction is repulsive, we only
include the contribution from the open states, like in \eq\nr{barS1_2}.
The octet Sommerfeld factor reads~\cite{fadin}
\be
  \So = \frac{ \Xo } { e ^{  \Xo } -1 }
  \;, \quad    
  \Xo  = \Bigl( \frac{\Nc}{2} - \CF \Bigr) \frac{ g ^ 2 } { 4 v }
  \;. \label{S8}
\ee
Numerical results are displayed in \fig\ref{fig:pert}, showing
a modest suppression.

%
\section{Lattice analysis}
\la{se:lattice}

We have measured the observables defined in
\eqs\nr{barS1_final}--\nr{barS8_final} 
using lattice NRQCD. The gauge configuration 
ensemble corresponds to $\Nf = 2+1$ dynamical quark flavors at
a temperature of a few hundred MeV. 
In this exploratory study, a single lattice spacing
and a single spatial volume were employed. The same thermal
gauge configurations have previously been used for 
studying in-medium modifications of bottomonium spectral
functions~\cite{lat1b} and the electric conductivity 
of the quark-gluon plasma~\cite{lat1d}. 

The configurations used correspond to an
anisotropic lattice setup, with the spatial lattice
spacing $a_s$ and the temporal one $a_t$ related through 
$a_s/a_t \approx 3.5$. 
The tuning of the anisotropy parameter 
and the vacuum properties 
of the lattice action were studied
by the Hadron Spectrum Collaboration~\cite{lat0a,lat0b}, yielding
$a_s = 0.1227(8)$~fm, $M_\pi \simeq 400$~MeV,
$M_K \simeq 500$~MeV. The critical temperature 
was estimated from the behaviour of the Polyakov loop expectation
value~\cite{lat1a}, with the result $\Tc = 185(4)$~MeV. 
The heavy quark mass for the physical bottom case, 
$a_s M^{ }_\rmi{kin}  = 2.92$, was tuned 
by extracting a kinetic mass from 
dispersion relations and matching it onto 
the spin-averaged mass of the $\eta_b$ and $\Upsilon$
mesons. Further details on the lattice setup
can be found in ref.~\cite{lat1b}.

It is important to note that the contribution of the bare heavy quark ``rest
mass'' drops out in the ratios in \eqs\nr{barS1_final} and
\nr{barS8_final}, and radiatively generated self-energy divergences
drop out as well. Therefore we can leave out the bare rest mass from
the equation of motion defining the heavy quark propagator, which
facilitates the measurement.
The form of the NRQCD propagator is specified in appendix~A.

%
\begin{table}[t]

\small{
\begin{center}
\begin{tabular}{lllllllll}
 \hline \\[-3mm]
 $\Nt$ &
 $ T / \Tc $ &
 $ a_s M^{ }_\rmi{kin}$ &   
 $ M^{ }_\rmi{kin} / T$ &  
 $10^4\, P^{ }_1$ &
 $10^5\, P^{ }_2$ &
 $10^6\, P^{ }_3$ &
 $\bar{S}^{ }_1$ &
 $\bar{S}^{ }_8$   \\[3mm]
 \hline 
 32 & 0.95 & 2.92 & 26.7 
 & 0.75(7)& 0.938(5)& 1.06(1) & 1690(410) & 3.6(15)
 \\ 
 28 & 1.09 & 2.92 & 23.4 
  & 3.0(1) & 2.78(1) & 3.19(3) & 301(35) & 1.2(3)     
 \\ 
 24 & 1.27 & 2.92 & 20.0 
  & 8.0(3) & 8.64(3) & 10.3(1) & 134(13) & 1.2(2)     
 \\ 
 20 & 1.52 & 2.92 & 16.7 
  & 24.4(5) & 28.5(1) & 36.1(4) & 48(3) & 0.83(6)     
 \\ 
 16 & 1.90 & 2.92 & 13.3 
  & 68.9(9) & 102.2(3) & 147(1) & 21.6(6) & 0.80(2)    
 \\ 
 \hline 
 32 & 0.95 & 1.50 & 13.7 
  & 1.31(7)& 1.308(8)& 1.51(1) & 758(128) & 3.7(5)     
 \\
 28 & 1.09 & 1.50 & 12.0 
  & 4.6(1)& 3.04(2)& 3.62(3) & 142(15) & 1.25(9)     
 \\ 
 24 & 1.27 & 1.50 & 10.3 
  & 10.4(2) & 7.38(4)& 9.21(8) & 69(5) & 1.05(5)    
 \\ 
 20 & 1.52 & 1.50 & 8.57 
  & 25.7(3) & 19.7(1) & 27.2(2) & 30(2) & 0.89(3)    
 \\ 
 16 & 1.90 & 1.50 & 6.86 
  & 61.9(6) & 60.1(3) & 95.8(8) & 15.7(3) & 0.85(1)    
 \\ 
 \hline 
 16 & 1.90 & 2.00 & 9.14 
  & 57.6(6) & 55.7(2) & 86.4(8) & 16.8(4) & 0.83(1)     
 \\
 16 & 1.90 & 2.50 & 11.4 
  & 62.5(7) & 76.0(2) & 113(1) & 19.5(5) & 0.81(2)     
 \\ 
 \hline 
\end{tabular} 
\end{center}
}

\vspace*{3mm}

\caption[a]{\small
 Parameters and results of the unquenched $\Nf = 2+1$ flavour
 simulations, with
 $\Nt$ denoting the number of lattice points in the time direction. 
 The 4-volume in lattice units was fixed at $\Nt \times 24^3 $; 
 the lattice spacings are asymmetric, with $a_s / a_\tau \approx 3.5$.
 The same configurations ($10^3$ per parameter set) were previously used 
 e.g.\ in refs.~\cite{lat1b,lat1d}. Errors are statistical only; 
 for $\bar{S}^{ }_1$ and $\bar{S}^{ }_8$ they were obtained with
 a jackknife analysis. The results of this table
 are illustrated in \fig\ref{fig:unquenched}. 
 }
\label{table:unquenched}
\end{table}
%

%
\begin{table}[t]

\small{
\begin{center}
\begin{tabular}{lllllllll}
 \hline \\[-3mm]
 $\beta^{ }_0$ &
 $ T / \Tc $ &
 $ a_s M^{ }_\rmi{kin}$ &   
 $ M^{ }_\rmi{kin} / T $ & 
 $10^2\, P^{ }_1 $ &
 $10^3\, P^{ }_2 $ &
 $10^4\, P^{ }_3 $ &
 $\bar{S}^{ }_1$ &
 $\bar{S}^{ }_8$   \\[3mm]
 \hline 
 5.85 & 1.33 & 3.289 & 13.2 
  & $2.5(1)$ & $6.26(4) $ &
  $11.4(5) $ & 10.3(7) & 0.81(7)  \\ 
 5.90 & 1.47 & 2.988 & 12.0 
  & $2.06(8)$ & $4.19(3)$ & 
  $7.9(3)  $ & 9.9(6) & 0.86(5) \\ 
 5.95 & 1.62 & 2.724 & 10.9 
  & $2.17(6)$ & $3.07(2)$ & 
  $6.9(3)  $ & 6.6(2) & 0.83(1)  \\ 
 6.00 & 1.77 & 2.498 & 10.0 
  & $2.01(5)$ & $2.20(1)$ & 
  $5.6(2)  $ & 5.4(3) & 0.86(2)  \\ 
 6.05 & 1.93 & 2.302 & 9.21 
  & $1.81(4)$ & $1.64(1)$ & 
  $4.2(1)  $ & 5.0(2) & 0.83(2)  \\ 
 \hline 
\end{tabular} 
\end{center}
}

\vspace*{3mm}

\caption[a]{\small Parameters and results of the quenched simulations,
  with $\beta^{ }_0$ denoting the coefficient of the Wilson plaquette
  term.  The 4-volume in lattice units was fixed at $4 \times 8 ^3 $.
  Conversions to units of $\Tc$ are based on ref.~\cite{betac}. We
  show these small-scale isotropic ($a_s / a_\tau = 1$) lattice
  results (from 200 gauge configurations)
  in order to permit for a rapid crosscheck of the
  measurement algorithm. } 
\label{table:quenched}
\end{table}
%

The main results of our lattice study are given in
table~\ref{table:unquenched} and in \fig\ref{fig:unquenched}. We note
that in
addition to our main $\Nf = 2 + 1$ lattice calculation, we have also
carried out tests on quenched lattice gauge configurations with a
symmetric lattice spacing on a small lattice volume 
$4 \times 8^3$. For completeness the result of quenched test is shown in
table~\ref{table:quenched}. 


\begin{figure}[t]

\hspace*{-0.1cm}
 \centerline{%
  \epsfxsize=7.5cm\epsfbox{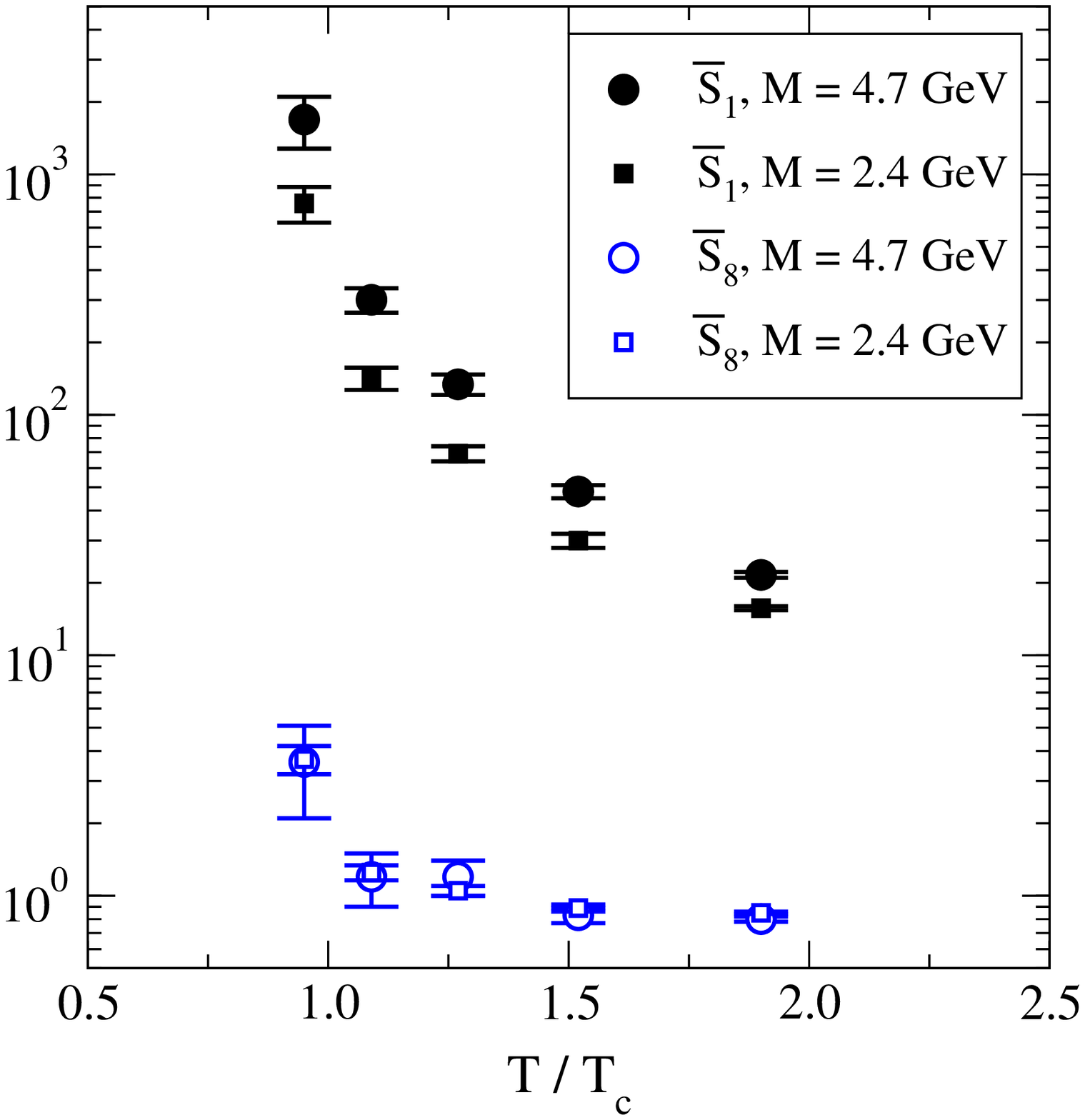}%
  \hspace{0.1cm}
  \epsfxsize=7.5cm\epsfbox{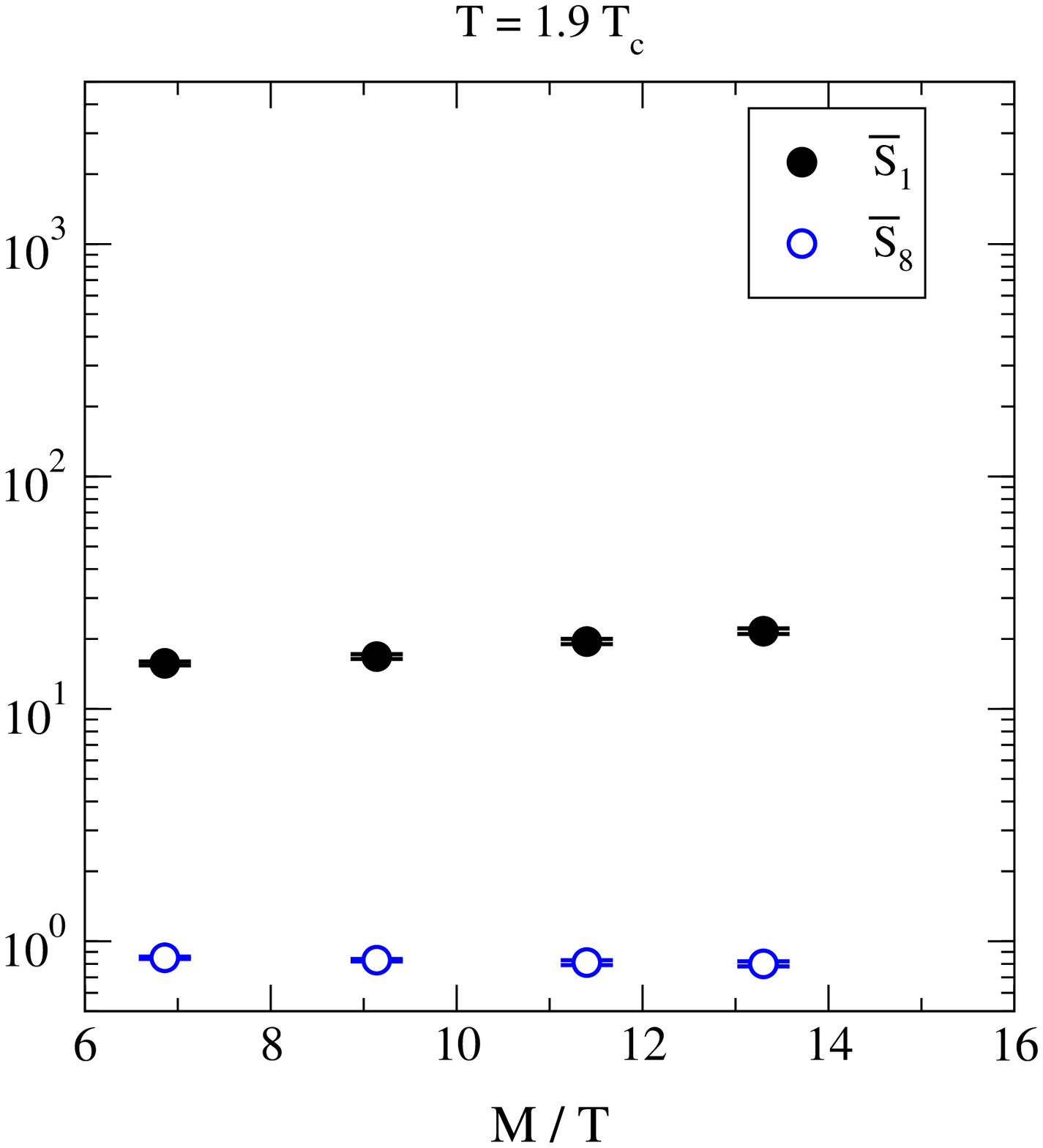}
 }

\caption[a]{\small
 Lattice NRQCD estimates of the thermal Sommerfeld factors. 
 Left: Results for fixed $M$ as a function of $T/\Tc$, 
 with $\Tc \simeq 185$~MeV. Right: Results for a fixed 
 $T/\Tc \approx 1.9$ as a function of $M/T$. The parameters cannot be chosen 
 exactly the same as in \fig\ref{fig:pert} because the lattice ensemble
 corresponds to an unphysical value of $M_\pi$ and does not represent
 an infinite-volume and continuum extrapolation. Nevertheless a reasonable
 qualitative agreement with \fig\ref{fig:pert} can be observed.   
}

\la{fig:unquenched}
\end{figure}

Comparing our $\Nf = 2+ 1$ lattice results 
in \fig\ref{fig:unquenched} with the perturbative
results in \fig\ref{fig:pert}, the following conclusions can be drawn:

\bi

\item[(i)] 
The lattice and perturbative results for $\bar{S}^{ }_8$ are
of similar magnitudes, and at $T \gsim 1.5 \Tc$ both are somewhat
below unity. 
At smaller $T/\Tc$ the central values of the lattice results are larger, 
however the 
observable in \eq\nr{barS8_final} involves a subtraction in the 
numerator and thus a large cancellation at $T \lsim 1.5 \Tc$. 
This is likely to lead to systematic uncertainties
from lattice artifacts which may be 
much larger than the statistical ones. 
  
\item[(ii)] 
The lattice results for $\bar{S}^{ }_1$ can exceed $10^3$, 
whereas the perturbative ones
according to the prescription used in the literature, 
which we have denoted by $\bar{S}^{ }_{1,\rmi{open}}$
(cf.\ \eq\nr{barS1_2}), never exceed 10. Even though the 
perturbative result may be expected to have $\sim 50$\%
uncertainties (see below), such a discrepancy of two orders
of magnitude cannot be accounted for by conceivable 
perturbative uncertainties. 

\item[(iii)] 
The lattice results for $\bar{S}^{ }_1$ are in qualitative agreement with 
the perturbative ones including bound state contributions, which we 
have denoted by $\bar{S}^{ }_{1,\rmi{open+bound}}$. Certain differences
do remain: the mass dependence at fixed $T$ is more rapid in the 
perturbative results (cf.\ right panels of 
\figs\ref{fig:pert} and \ref{fig:unquenched}), and the temperature scales 
are somewhat shifted (cf.\ left panels of 
\figs\ref{fig:pert} and \ref{fig:unquenched}). 
We stress, however, that
the vertical axes of the plots are logarithmic; 
without the inclusion of bound state
contributions, there would be differences of up to two orders of 
magnitude. With the inclusion of bound states, orders of magnitude can 
be accounted for. For the remaining discrepancies there are many 
likely explanations. First of all our leading-order 
perturbative computation has uncertainties on the $\sim 50$\% level.
Amongst others, the perturbative predictions
are sensitive to the parameter values used, in particular the scale
chosen for the gauge coupling and the scheme chosen for the quark mass; 
in the absence of an actual higher-order
computation of $\bar{S}^{ }_1$, the choices made
amount to {\em ad hoc} recipes. Second, the lattice results correspond
to quark masses larger than in the continuum computation, which implies
that physical scales cannot be compared unambiguously (this is reflected
by the different values of $\Tc/$MeV). Third, the
lattice results have no continuum limit and take place in a finite
volume, $V\approx\;$($2.9\,\mbox{fm}$)$^3$. 
It would be interesting
to carry out a refined lattice analysis and a systematic higher-order
perturbative computation in order to see if the discrepancies 
get thereby reduced. 

\ei

%
\section{Discussion and outlook}
\la{se:concl}

We have presented a power-counting argument showing that bound states 
(if they exist) 
dominate the thermal co-annihilation rate of kinetically equilibrated
non-relativistic particles at low temperatures
(discussion around \eq\nr{robust}),  
as well as a theoretical derivation of basic formulae 
(\eqs\nr{master}, \nr{barS1_final}, \nr{barS8_final} 
and \nr{rho_open_bound}) which
permit for the inclusion of bound state contributions in  
perturbative (\se\ref{se:pert}) or
non-perturbative (\se\ref{se:lattice}) studies. 
The perturbative side can be reduced to the determination of
a spectral function, i.e.\ 
the imaginary part of a particular Green's function, 
with a thermal potential incorporating
the effects of Debye screening and soft $2\to 2$ scatterings 
of the co-annihilating particles with light plasma constituents. 
The results of both perturbative and non-perturbative 
computations can be parametrized by generalizations 
of the so-called thermal Sommerfeld factors. 

Even though our study was inspired by the co-annihilation 
rate of weakly interacting non-relativistic particles in cosmology, 
which determines their freeze-out temperature, we applied the formalism
to obtain a non-perturbative estimate of the heavy quark chemical 
equilibration rate in QCD, relevant 
for heavy ion collision experiments. 
Our numerical investigation confirmed
the existence of Sommerfeld enhancement in the case of attractive 
interactions as well as a reduction (at $T \gsim 1.5\Tc$) 
in the case of repulsive
interactions (cf.\ \fig\ref{fig:unquenched}). 
At $T \gsim 1.5\Tc$ the repulsive case shows 
reasonable agreement with perturbation theory
(cf.\ \fig\ref{fig:pert}). 

In the attractive case, 
corresponding to the factor $\bar{S}^{ }_1$, 
we found a much larger Sommerfeld enhancement than 
predicted by the perturbative formula sometimes used in literature 
(cf.\ \eq\nr{barS1_2}). In accordance with the power-counting 
argument around \eq\nr{robust} 
we believe that the difference originates
from bound state contributions, 
which were omitted in \eq\nr{barS1_2}.\footnote{%
 Bound state contributions have been included in thermal 
 QCD literature in other contexts, for instance in studies of 
 the leptonic decays of heavy quark-antiquark pairs. However 
 these processes are too slow by many orders of magnitude to 
 contribute to heavy quark chemical equilibration. 
 }
Due to their smaller mass, bound states appear with a less
suppressed Boltzmann weight in the thermal ensemble than scattering states. 
Note also that bound state formation is a relatively speaking
fast process, without any exponential suppression factors; 
thermal friction, caused by $2\to2$ scatterings with light
plasma constituents, lets an open $q\bar{q}$ pair lose energy until
it appears with its proper thermal weight. Including bound state 
decays through a numerical determination of the imaginary 
part of a thermal Green's function, $\rho^{ }_{q\bar{q}}(E,k)$,  
we did find qualitative agreement between perturbative and 
lattice results (cf.\ \figs\ref{fig:pert} and \ref{fig:unquenched}).

It may be asked to what extent our findings could be relevant
for the weakly interacting cases considered in cosmology. First of all 
we note that typical 
parameter values playing a role in cosmology 
are $\alpha_\rmi{w} \sqrt{M/T} \sim 0.04 \sqrt{20} \sim 0.2$, 
whereas we have 
simulated $\alphas \sqrt{M/T} \sim 0.25 \sqrt{(2.4 ... 4.7)/(0.2 ... 0.4)} 
\sim 0.6 ... 1.2$.
This means that we have been deeper in the enhancement regime than is 
typical for cosmology. That said, there are cosmological models 
in which bound state formation has been demonstrated in vacuum  
(cf.\ e.g.\ refs.~\cite{wimpo1,wimpo2,wimpo3,wimpo4}).
It might be interesting to use our formalism to study whether
or not bound states could affect the thermal 
freeze-out process in these cases. Another direction worth a look
are models based on strongly interacting dark sectors 
(cf.\ e.g.\ refs.~\cite{dark1,dark2,dark3,dark35} and references therein).

Let us finally return from cosmology to hot QCD.  Inserting the values
of the absorptive parts of the 4-quark coefficients from
ref.~\cite{bodwin} and assuming that the octet Sommerfeld factors are
spin-independent, the heavy quark chemical equilibration rate
evaluates to~\cite{pert} \be \Gamma^{ }_\rmi{chem} \; \simeq \;
\frac{8\pi\alphas^2}{3 M^2} \left ( \frac{ M T } { 2 \pi } \right ) ^{
  3/2 } e ^{ - M/T } \biggl[ \frac{\bar{S}^{ }_1}{3} + \biggl( \fr56 +
  \Nf \biggr) \bar{S}^{ }_8 \biggr] \;. \la{pheno} \ee Inserting the
highest temperature reached in current heavy-ion collisions, $T\sim
400$~MeV, a charm quark mass $M \sim 1.5$~GeV, a running coupling
$\alphas\sim 0.25$, and values $\bar{S}_1 \sim 15$, $\bar{S}_8 \sim
0.8$ as estimated in the current study for $\Nf = 3$, we get
$\Gamma^{-1}_\rmi{chem} \sim 150$~fm/c. Therefore heavy quark chemical
equilibration is unlikely to take place within the lifetime $\sim
10$~fm/c of the current generation of heavy ion collision
experiments. This can be compared with their kinetic equilibration
time scale which could be as small as $\sim 1$~fm/c in the case of
charm quarks~\cite{latt_c,kappa}. It may be noted, however, 
that $\Gamma^{ }_\rmi{chem}$ (\eq\nr{pheno}) 
changes rapidly with temperature. Already 
a modest increase, in the ballpark of the planned Future
Circular Collider (FCC) heavy ion program~\cite{fcc}, could therefore 
yield chemically equilibrated charm quarks.



%
\section*{Acknowledgements}

We thank the FASTSUM collaboration for providing the 
gauge configurations used in our unquenched measurements.
M.L.\ thanks D.~B\"odeker for collaboration
at initial stages of this project, and M.~Garny for a helpful discussion. 
S.K.\ thanks AEC of the University of Bern for hospitality and support
during the completion of this work. S.K.\ was supported by the
National Research Foundation of Korea under
grant No.\ 2015R1A2A2A01005916
funded by the Korean government (MEST).
M.L.\ was supported by the Swiss National Science Foundation
(SNF) under grant 200020-155935. 

%
\appendix
\renewcommand{\thesection}{Appendix~\Alph{section}}
\renewcommand{\thesubsection}{\Alph{section}.\arabic{subsection}}
\renewcommand{\theequation}{\Alph{section}.\arabic{equation}}

%
\section{Heavy quark propagators}

For completeness we 
reiterate here the non-perturbative definitions of heavy quark 
and antiquark propagators within the NRQCD framework. Denoting 
\ba
 G^\theta_{\alpha\gamma;ij}(\tau_2,\vc{x};\tau_1,\vc{y}) & \equiv & 
 \Bigl\langle 
 \theta^{ }_{\alpha i}( \tau_2,\vc{x} )\, 
 \theta^*_{\gamma j} (\tau_1,\vc{y} ) 
 \Bigr\rangle
 \;, \la{prop_theta} \\
 G^\chi_{\alpha\gamma;ij}(\tau_2,\vc{x};\tau_1,\vc{y}) & \equiv & 
 \Bigl\langle 
 \chi^{ }_{\alpha i}( \tau_2,\vc{x} )\, 
 \chi^*_{\gamma j} (\tau_1,\vc{y} ) 
 \Bigr\rangle
 \;, \la{prop_chi}
\ea
the fields ``naturally'' propagate into the regions $\tau_2 > \tau_1$ and
$\tau_1 > \tau_2$, respectively, given that 
$\theta^\dagger$ and $\chi$ correspond to creation operators
for quarks and antiquarks. 
Specifically, in continuum notation and in a given gauge field background, 
\be
 \biggl( D^{ }_{\tau_2} + M_\rmi{rest} - \frac{\vec{D}^2 
 }{2 M_\rmi{kin}}
 + \ldots 
 \biggr)
 G^\theta_{ } (\tau_2,\vc{x} ; \tau_1,\vc{y}) = 0 
 \;, \quad \tau_2 > \tau_1 
 \;, \la{propG}
\ee
with the initial condition
\be
 G^\theta_{ } (\tau_1^+,\vc{x};\tau_1,\vc{y}) = 
 \unit^{ }_{2\Nc \times 2\Nc} \, \delta^{(3)}(\vc{x}-\vc{y})
 \;.
\ee
The boundary condition 
\be
 G^\theta_{ } (0,\vc{x};\tau_1,\vc{y}) \equiv 
 - 
 G^\theta_{ } (\beta,\vc{x};\tau_1,\vc{y})
 \la{bcG}
\ee
allows us to go into the region $\tau_2 < \tau_1$
(cf.\ \fig\ref{fig:Gcirc}). 
The propagator for $\chi$ is obtained from 
\be
 G^\chi(\tau_2,\vc{x};\tau_1,\vc{y}) = 
 - \bigl[ G^\theta(\tau_1,\vc{y};\tau_2,\vc{x}) \bigr]^\dagger
 \;. \la{Gchi}
\ee

We stress that, as mentioned in the text, the contribution of 
the rest mass, $M^{ }_\rmi{rest}$, 
appearing linearly in \eq\nr{propG}, drops out in our
final observables, \eqs\nr{barS1_final} and \nr{barS8_final}. 
Therefore this term can be omitted from the practical
computations.

On the lattice 
the non-relativistic quark propagator is calculated from 
($\cdot\equiv 0,\vc{0}$)
\begin{eqnarray}
G^\theta (0,{\mathbf x};\cdot) & = & 
  \frac{\delta_{{\mathbf x}, \vc{0}}}{a_s^3}  \;,  \\
G^\theta (a_t,{\mathbf x};\cdot) & = &
  \left(1 - \frac{a_t \mathcal{H}^{ }_0}{2n}\right)^n
U_0^\dagger(0,{\mathbf x})
  \left(1 - \frac{a_t \mathcal{H}^{ }_0}{2n}\right)^n 
G^\theta(0,{\mathbf x};\cdot) \;, \\
G^\theta ( \tau+a_t,{\mathbf x};\cdot) & = &
  \left(1 - \frac{a_t \mathcal{H}^{ }_0}{2n}\right)^n
 U_0^\dagger(\tau,{\mathbf x}) 
 \left(1 - \frac{a_t \mathcal{H}^{ }_0}{2n}\right)^n 
 \bigl( 1 - a_t\,\delta \mathcal{H}  \bigr)  
 \, G^\theta (\tau,{\mathbf x};\cdot) \;, \hspace*{5mm} 
\end{eqnarray}
where $U_0^{ }$ is a time-direction gauge link.
The lowest-order Hamiltonian reads
\be
 \mathcal{H}^{ }_0 = - \frac{\Delta^{(2)}}{2M^{ }_\rmi{kin}}
 \;,
\ee
where $\Delta^{(2)}$ is a discretized gauge Laplacian. 
The higher order correction is
\begin{eqnarray}
\delta \mathcal{H} & = &  - \frac{(\Delta^{(2)})^2}{8 M^3_\rmi{kin}} +
\frac{ig^{ }_0\, 
 ({\bf \nabla}\cdot {\bf E} - {\bf E}\cdot {\bf \nabla})
 }{8 M^2_\rmi{kin}}  
  - \frac{g^{ }_0\,
 {\bf \sigma} \cdot 
  ({\bf \nabla}\times {\bf E} - {\bf E}\times {\bf \nabla})  
   }{8 M^2_\rmi{kin}}   \nn 
&&
 \; - \, 
 \frac{g^{ }_0\, {\bf \sigma} \cdot {\bf B}
  }{2 M^{ }_\rmi{kin}} 
 + \frac{a_s^2 \Delta^{(4)}}{24 M^{ }_\rmi{kin}}
 - \frac{a_t (\Delta^{(2)})^2}{16 n M^2_\rmi{kin}}
 \;, 
\label{eq:deltaH}
\end{eqnarray}
where $g^{ }_0$ is the bare gauge coupling, and 
unspecified notation is explained
in ref.~\cite{lat1b}. 
The parameter $n$, which stabilizes the high-momentum behaviour of
the propagator, is set to 1 for 
$a_s M^{ }_\rmi{kin} = 2.92$ and to 3 for $a_s M^{ }_\rmi{kin} = 1.5$.
The electric and magnetic fields appearing 
in \eq\nr{eq:deltaH} were implemented in 
a tadpole-improved~\cite{Lepage:1992xa} form, with 
the improvement factor 
$u_s = 0.7336$ for the spatial link and $u_\tau = 1.0$ for the time
link~\cite{lat0a}. 

%

\end{document}